\documentclass[12pt,twocolumn]{aastex62}

\usepackage{graphicx}
\usepackage{apjfonts}
\usepackage{epsfig}
\usepackage{natbib}

\newcommand\sig{$\sigma$}
\newcommand\mstar{$M^*$}
\newcommand\kms{km~s$^{-1}$}
\newcommand\hbeta{H$\beta$}
\newcommand\feh{[Fe/H]}
\newcommand\alp{$\alpha$}
\newcommand\afe{[\alp/Fe]}

\newcommand\feave{$\langle {\rm Fe} \rangle$}
\newcommand\atd{ATLAS$^{\rm 3D}$}
\newcommand\msun{$M_{\odot}$}
\newcommand\hst{\emph{HST}}

\newcommand{\aref}[1]{\mbox{\hyperref[#1]{Appendix~\ref{#1}}}}

\begin{document}

\title{The MASSIVE Survey - XII Connecting Stellar Populations of Early-Type Galaxies to Kinematics and Environment}

\correspondingauthor{Jenny E. Greene}
\email{jgreene@astro.princeton.edu}

\author{Jenny E. Greene} 
\affil{Department of Astrophysical Sciences, Princeton University, Princeton, NJ 08544, USA}
\author{Melanie Veale}
\affil{Department of Astronomy, University of California, Berkeley, CA 94720, USA}
\author{Chung-Pei Ma}
\affil{Department of Astronomy, University of California, Berkeley, CA 94720, USA}
\affil{Department of Physics, University of California, Berkeley, CA 94720, USA}
\author{Jens Thomas}
\affil{Max Plank-Institute for Extraterrestrial Physics, Giessenbachstr. 1, D-85741 Garching, Germany}
\author{Matthew E. Quenneville}
\affil{Department of Physics, University of California, Berkeley, CA 94720, USA}
\author{John P. Blakeslee}
\affil{Gemini Observatory, Casilla 603, La Serena, Chile}
\author{Jonelle L. Walsh}
\affil{George P. and Cynthia Woods Mitchell Institute for Fundamental Physics and Astronomy, and Department of Physics and Astronomy, \\Texas A\&M University, College Station, TX 77843, USA}
\author{Andrew Goulding}
\affil{Department of Astrophysical Sciences, Princeton University, Princeton, NJ 08544, USA}
\author{Jennifer Ito}
\affil{Department of Astronomy, University of California, Berkeley, CA 94720, USA}

\begin{abstract}
We measure the stellar populations as a function of radius for 90
early-type galaxies (ETGs) in the MASSIVE survey, a volume-limited
integral-field spectroscopic (IFS) galaxy survey targeting all
northern-sky ETGs with absolute $K$-band magnitude $M_K < -25.3$ mag,
or stellar mass $M_* \ga 4 \times 10^{11} M_\odot$, within 108 Mpc.
We are able to measure reliable stellar population parameters for
individual galaxies out to $10-20$ kpc ($1-3 R_e$) depending on the
galaxy.  Focusing on $\sim R_e$ ($\sim 10$ kpc), we find
significant correlations between the abundance ratios, \sig, and
\mstar\ at large radius, but we also find that the abundance ratios 
saturate in the highest-mass bin. We see a strong correlation between 
the kurtosis of the line of sight velocity distribution ($h4$) and 
the stellar population parameters beyond $R_e$. Galaxies with higher radial 
anisotropy appear to be older, with metal-poorer stars and enhanced 
[$\alpha$/Fe]. We suggest that the higher radial anisotropy may derive 
from more accretion of small satellites. Finally, we see some evidence for
correlations between environmental metrics (measured locally and on
$>5$~Mpc scales) and the stellar populations, as expected if 
satellites are quenched earlier in denser 
environments.
\end{abstract}




\section{Introduction}
\label{sec:introduction}

The late-time assembly history of massive early-type galaxies remains
a topic of ongoing interest. At early times, typical quiescent
galaxies were quite compact
\citep[e.g.,][]{vandokkumetal2008,vanderweletal2008}. At the present
time, massive early-type galaxies typically have extended
envelopes
\citep[e.g.,][]{schombert1986,kormendyetal2009,huangetal2013,huangetal2018a}. To
some degree, the larger sizes reflect that at fixed mass, larger galaxies join the
red sequence at later times \citep[e.g.,][]{newmanetal2012}, but most
massive galaxies also likely build up their outskirts through the
accretion of smaller satellites that dissolve at large radius
\citep[e.g.,][]{naabetal2009,bezansonetal2009}. Cosmological
hydrodynamical simulations from numerous groups have shown that there
are two phases to the build-up of stellar mass in massive early-type
galaxies, with a gas-rich phase forming a compact core at early times
($z \approx 2$) followed by dissipationless merging at late times
\citep[e.g.,][]{oseretal2010,rodriguez-gomezetal2016,wellonsetal2016}. From
an observational perspective, it is still unclear what fraction of the
size growth can be explained by minor mergers as opposed to the
quenching of larger galaxies
\citep[e.g.,][]{valentinuzzietal2010,newmanetal2012,barroetal2013}.
Observations of the stellar populations and kinematics of local
galaxies at large radius may provide complementary insights into the
assembly history of massive galaxies.

In the simulations, the fraction of accreted (``ex-situ'') stars
rises with both halo mass and stellar mass
\citep[e.g.,][]{oseretal2010,rodriguez-gomezetal2016}. Photometric
observations of local massive early-type galaxies have presented some
confirmation for a two-phase formation scenario
\citep{huangetal2013,dsouzaetal2014,ohetal2017}, while
\citet{huangetal2018a,huangetal2018b} present empirical evidence that
the ex-situ fraction correlates both with stellar and halo mass.
Spectroscopic observations should provide complementary constraints on
the assembly history of massive elliptical galaxies. More
specifically, the radial gradients in stellar populations should
encode the build-up of stellar mass, particularly if we can reach to
large radius.  It is very challenging to obtain high signal-to-noise
(S/N) observations at large galactocentric radius, and the bulk of papers
looking at radial gradients in stellar populations have worked within
$R_e$
\citep[e.g.,][]{spinradtaylor1971,mehlertetal2003,annibalietal2007,
  spolaoretal2010,jimmyetal2013,mcdermidetal2015,goddardetal2017}.  In
recent years, integral-field spectroscopy (IFS) has enabled stellar
population measurements at large radius
\citep[e.g.,][]{greeneetal2013,scottetal2013,greeneetal2015,mcdermidetal2015,goddardetal2017,boardmanetal2017,
  baroneetal2018,vandesandeetal2018}.

Here we focus on the most massive galaxies in the
present-day universe using the MASSIVE survey \citep{maetal2014}. We
have gathered integral-field data for $90$ MASSIVE
galaxies and measured their stellar kinematics including higher-order
moments \citep{vealeetal2017a,vealeetal2017b}, stellar velocity
dispersion profiles \citep{vealeetal2018}, and kinemetry
\citep{eneetal2018}. In addition, we have \hst/WFC3 imaging for 30 of
the galaxies \citep{goullaudetal2018}, and uniform CFHT $K-$band
imaging for nearly all objects.

\begin{figure*}
\begin{center}
\includegraphics[width=0.65\columnwidth]{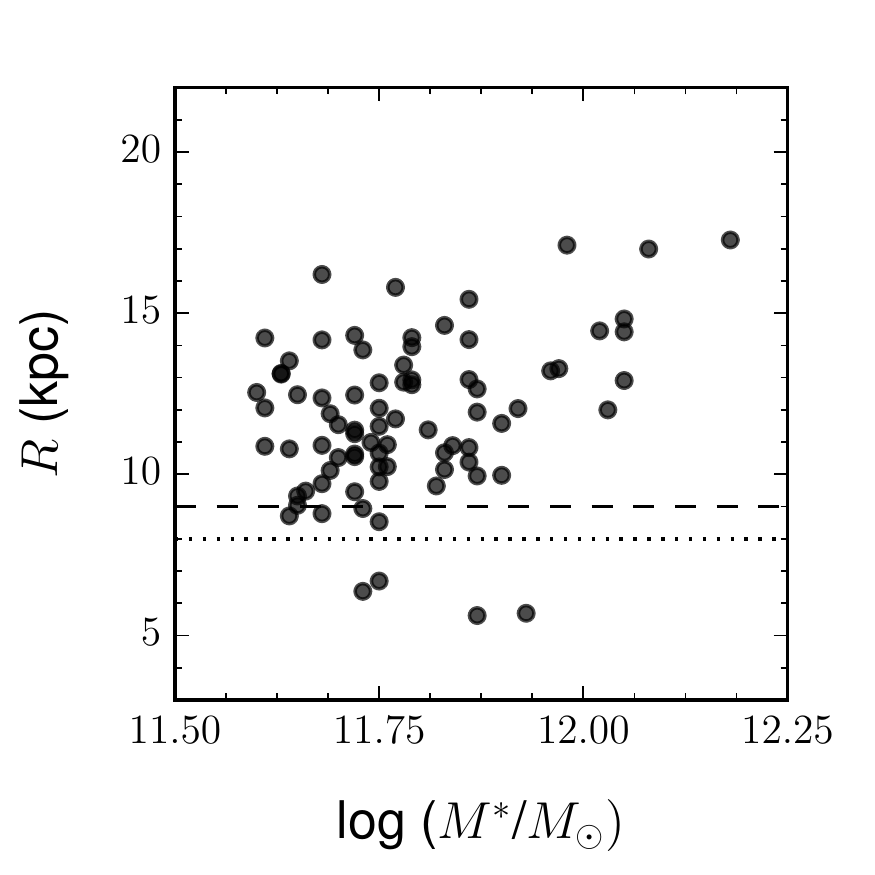}
\includegraphics[width=0.65\columnwidth]{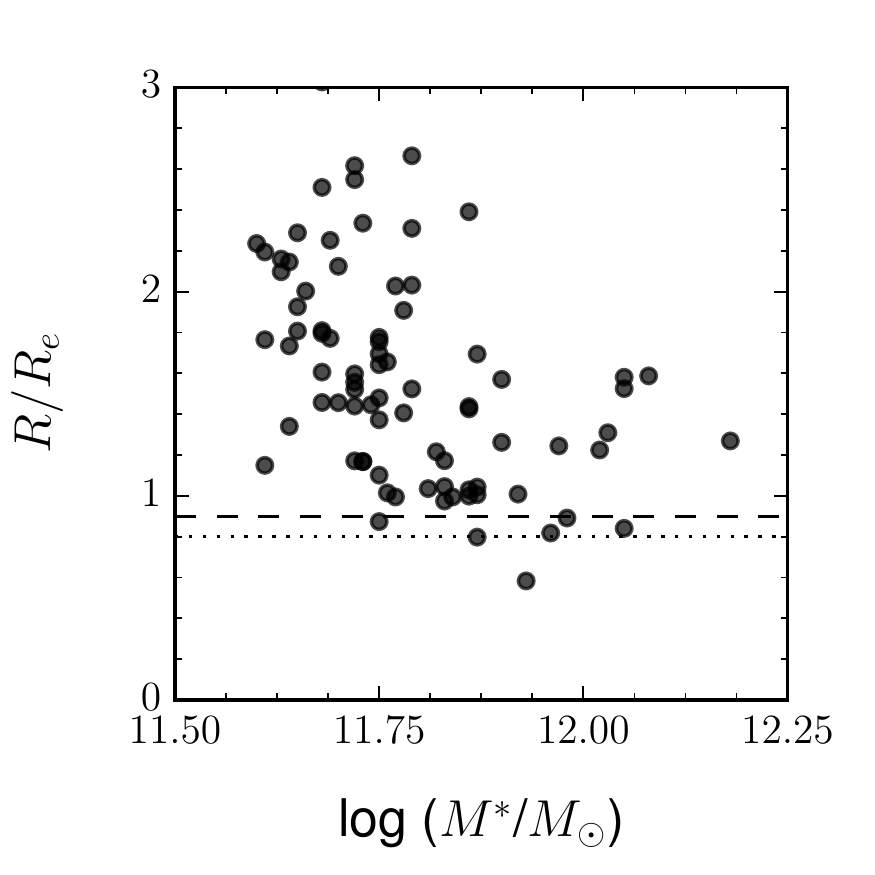}
\includegraphics[width=0.65\columnwidth]{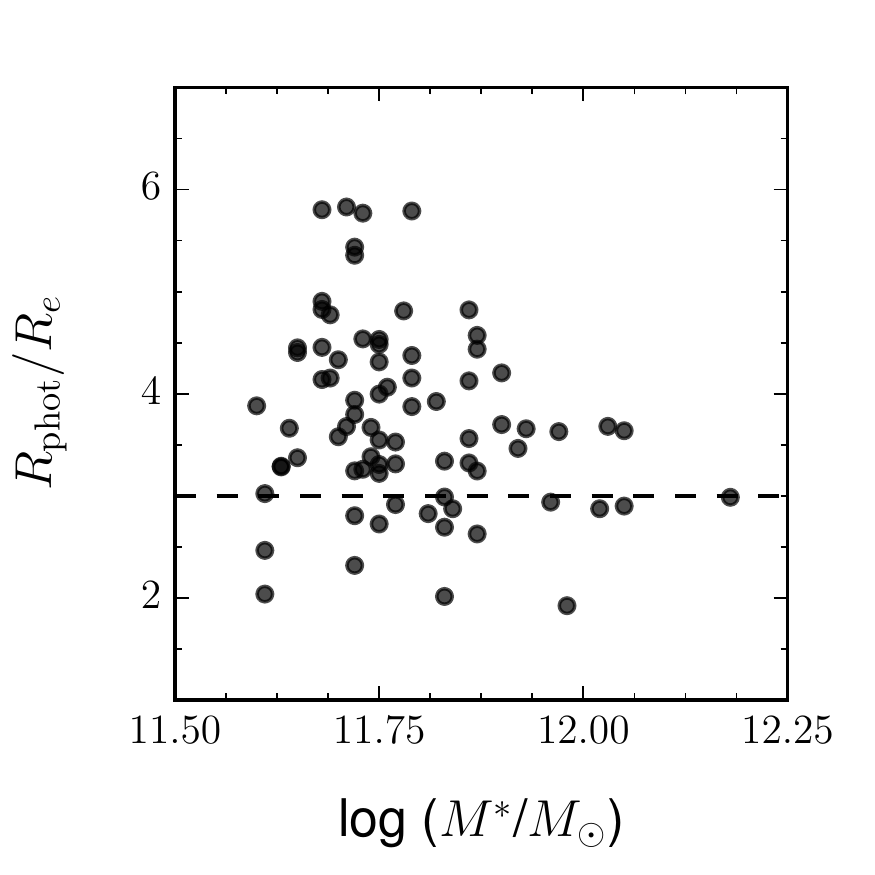}
\end{center}
\caption{{\it Left}: 
The maximum radial coverage of our sample as a function of stellar mass. 
We take a fiducial ``outer'' 
measurement at 9 kpc (dashed line) by weighting the measurements by proximity 
to this radius. Only galaxies with coverage to 8 kpc (dotted line) are 
included, which allows us to keep nearly all of the galaxies.
{\it Middle}: Same as left, but taking a radius of $0.9~R_e$.
{\it Right}: Radial coverage of the photometry, in $R_e$ units.}
\label{fig:radial}
\end{figure*}

\section{Galaxy Sample}
\label{sec:sample}

MASSIVE is a volume-limited survey of 116 galaxies \citep[see details
  in][]{vealeetal2017b} in the Northern hemisphere within $D<108$~Mpc
(i.e., to the distance of the Coma cluster), with $K-$band magnitudes
$M_K<-25.3$~mag (roughly \mstar$>10^{11.5}$~\msun). Details of the
target selection are described in \citet{maetal2014}. Briefly, our
targets are drawn from the Two Micron All-sky Survey
\citep[2MASS;][]{skrutskieetal2006} Extended Source Catalog
\citep[XSC;][]{jarrettetal2003} combined with distances from the high
density contrast (HDC) group catalog from \citet{crooketal2007},
supplemented by surface brightness fluctuations when available
\citep[][]{blakesleeetal2009,blakesleeetal2010,blakeslee2013} and with
the flow model of \citet{mouldetal2000} when needed.

We have completed integral-field spectroscopic (IFS) observations for
90 galaxies, including complete coverage of the 75 galaxies having
$M_K<-25.5$ mag. We use the Mitchell IFS at McDonald Observatory
\citep{hilletal2008}, which has a $107 \times 107$ arcsec$^2$
field-of-view, one-third of which is filled by 246 fibers of
4\arcsec\ diameter.  Each galaxy is observed at three dither positions
with twenty minute exposures, interspersed with ten minute sky
observations. The spectra span 3650-5850 \AA, with a spectral
resolution of $4.5$\AA\ full-width at half-maximum. The data are
reduced using the custom software Vaccine
\citep{adamsetal2011,murphyetal2011}. For more details see
\citet{greeneetal2015}.

A primary goal of the MASSIVE survey is to measure spatially resolved
stellar kinematics for dynamical modeling
\citep[e.g.,][]{thomasetal2016}.  To achieve a mean signal-to-noise
(S/N) of at least 20 pixel$^{-1}$, we group individual fibers into
spatial bins and co-add the spectra from fibers in a given bin into a
single spectrum.  The binning scheme is described in detail in
\citet{vealeetal2017a}, but we summarize the procedure here briefly
for completeness. Central fibers with S/N$>20$ are kept as their own
individual bins. Outside of the central regions, the fibers are
`folded' across the major axis to boost S/N. The fibers are then
grouped into annular bins, and each annulus is sub-divided into an
even number of angular bins. The radial extent of each bin is chosen
to achieve the target S/N of 20, subject to the constraint that the
aspect ratio of each bin, [0.5($R_{\rm outer}+R_{\rm inner} \Delta
\theta$)/[$R_{\rm outer}-R_{\rm inner}$], be less than or equal to
  1.5. The radial extent of the bins increase outwards until it is no
  longer possible to achieve S/N$>20$, at which point the remaining
  fibers are binned into two large radial bins.

Stellar kinematics are measured in each bin using pPXF
\citep{cappellariemsellem2004} and are presented in
\citet{vealeetal2017a}, including stellar velocity and velocity
dispersion, but also higher-order moments. The kinematics as a
function of environment are analyzed in \citet{vealeetal2017b} and
\citet{vealeetal2018}. Studies of the misalignments between kinematic
and photometric axes and local velocity features such as kinematically
distinct components are presented in \citet{eneetal2018}, where an
``unfolded'' binning scheme was used.

\subsection{Environment Measures}

We use two probes of the density field, as described in detail in
\citet{vealeetal2017b}. One useful measure of galaxy environment is
the large-scale density field surrounding a galaxy on the scale of
several Mpc. For this, we use the density field from
\citet{carricketal2015}, based on the 2M++ redshift catalog
\citep{lavauxhudson2011}. The 2M++ covers nearly the full sky to a
depth of $K=12.5$ mag, and includes 69,160 galaxies from 2MRS, the
Sloan Digital Sky Survey Data Release 7 (SDSS-DR7; Abazajian et
al. 2009), and the 6dF galaxy redshift survey Data Release 3
(6dFGRS-DR3; Jones et al. 2009). The Carrick et al.\ galaxy density
contrast $\delta_g \equiv (\rho_g-\overline{\rho})/\rho_g$ is the
luminosity-weighted density contrast smoothed with a 5.7 Mpc Gaussian
kernel. 

We also calculate $\nu_{10}$, a local galaxy density, by taking the
distance to the 10th nearest neighbour and measuring the luminosity
enclosed in this region. As discussed in detail in Appendix A of
\citet{vealeetal2017b}, we adopt an absolute magnitude limit of $M_K =
-23$ mag to identify the tenth nearest neighbor. Given the magnitude
limit of the 2MASS parent sample, we start to lose neighbors for
MASSIVE galaxies beyond 80 Mpc. However, Veale et al.\ estimate that
the $\nu_{10}$ values are impacted at a low level by this
incompleteness.

\subsection{Photometry}
\label{sec:sigma}

Our size and surface-brightness measurements come from
Canada-France-Hawai'i Telescope (CFHT) $K-$band imaging
(Quenneville et al.\ in prep). Elliptical isophotes are fitted to
each galaxy using ARCHANGEL \citep{schombert2007}.  Then, a curve of
growth is fit to the cumulative aperture luminosities as a function of
radius to yield the magnitude of the galaxy.  The half-light or
effective radius is determined as the radius enclosing 50\% of the
light determined from the curve of growth analysis. These radii are
measured along the major axis and have not been circularized. We find that 
the radii measured in this way are systematically larger than the 
2MASS radii by 20\% on average (see details in Quenneville et
al.\ in prep). The adopted effective radii for the 84 galaxies
with both CFHT photometry and Mitchell IFS data are included in Table
1.

We measure the outer slope of the surface brightness profile ($\Delta
\Sigma/\Delta \rm{log} R$), following \citet{pillepichetal2014} and
\citet{cooketal2016}, who show from Illustris simulations
\citep{vogelsbergeretal2013} that this slope tracks the ex-situ
fraction \citep[see also][]{huangetal2018a}. Using the ARCHANGEL
radial profiles, we fit surface brightness as a function of the log of
the radius in $R_e$ units, following Cook et al.\ We do not have
uniform coverage to $4 R_e$ for all targets (Figure \ref{fig:radial}),
so we explore how robust the outer slopes are as a function of radial
coverage.  We have 14 targets with coverage beyond $4 R_e$. For this
sample, we measure the slope from $2 R_e$ out to $3 R_e$, and $4
R_e$. We find that the slopes measured within the two
radial range are well correlated, although those measured within the more 
restricted radii are slightly smaller
on average by $0.2$. We thus adopt the slope measured between $2-3 R_e$ 
($\Delta \Sigma_{23}$ hereafter), as it can be uniformly measured
for 80 of the galaxies in our sample.

\section{Stellar Population Measurements}
\label{sec:stellarpops}

In this work, Lick indices are measured to trace the stellar
populations.  Lick indices were originally developed as a way to
extract stellar population information from spectra without flux
calibration 
\citep{burstein1985,faberetal1985,wortheyetal1992,trageretal1998}. Each
Lick index is a narrow region (typically $\sim 20$\AA\ wide) that is
mostly dominated by a single element. Isolating these regions allows
to study age (through the Balmer lines and particular \hbeta),
metallicity (\feh\ through Fe lines), and variable abundance ratios.
Of course in practice no window is impacted purely by a single
element; particularly at the velocity dispersion of our galaxies
(200-400 \kms) all indices are blends of multiple elements, as
summarized nicely in Table 1 in \citet{gravesschiavon2008}.

We adopt the code {\it lick\_ew} \citep{gravesschiavon2008} measures the Lick
indices and these are fed to {\it EZ\_Ages} \citep{gravesschiavon2008}
to convert the Lick indices to physical parameters (age, \feh,
[\alp/Fe]). The code uses pairs of indices, starting with \hbeta\ and
\feave, to iteratively solve for the age, abundance and abundance
ratios. The models use response functions from \citet{kornetal2005}
and synthesis models from \citet{schiavon2007}. 

We follow \citet{schiavon2007} and quote [Fe/H], which is directly
inferred from the Fe indices, rather than quote a total
metallicity. Total metallicity depends on oxygen (the most abundant
heavy element) and we do not measure oxygen directly. Instead, we
generally assume that [O/Fe] tracks [Mg/Fe].  In our standard runs, we
utilize the \alp-enhanced isochrone from \citet{salasnichetal2000} and
the default assumption that [O/Fe]$=0.5$ to match the \alp-enhanced
isochrone value. Since the development of {\it EZ\_Ages}, more
sophisticated modeling schemes have been developed that implicitly
solve for [O/Fe] using Lick indices
\citep{thomasetal2011,johanssonetal2012,wortheyetal2014}. As well,
full spectral modeling takes advantage of information in all of the
pixels and boosts the signal-to-noise of the final
determinations. This is our goal for the future. Since
\citet{conroyetal2014} show that Lick methods and full spectral
fitting recover the same basic ages, metallicities, and abundance
ratios from the same set of SDSS spectra, it is possible to
intercompare our results with both literatures.

\subsection{Radial Coverage}
\label{sec:stellarpops-outer}

With the advent of integral-field data, it has become increasingly
clear that the aperture used for stellar population measurements
impacts the final result; for instance \citet[][]{baroneetal2018} 
uncover trends with galaxy density and stellar populations within 
the effective radius that do not hold for ``central'' values. Thus, we wish to
exploit our IFS data by measuring all properties within multiple standardized
radii, both in physical and $R_e$ units. In this subsection, we first
describe the final sample that we adopt, after removing a few galaxies
with more limited radial coverage or S/N. We then discuss the primary
measurements that we use at different radii.  Ultimately, we will
adopt luminosity-weighted measurements measured at 
$0.9 R_e$ and 1.5 $R_e$, as well as fixed physical radii of 9~kpc and 
15~kpc, as motivated by the radial coverage of our data.

We start with 90 MASSIVE galaxies with Mitchell IFS data. We remove
five galaxies (NGC 910, NGC 1226, NGC 7052, UGC 3021, and UGC 10918)
from consideration for all large-radius tests, due to their very
limited radial coverage (in these cases due to poor observing 
conditions). We exclude them from analysis of both radial
gradients and large-radius measurements. These five galaxies do not
have any other properties in common (e.g., they are neither the most
or least massive, or largest or smallest galaxies).  We then examine
the radial coverage of our IFS observations for the remaining galaxies. In Figure
\ref{fig:radial}, we show the maximum radial extent of our binned data
in kpc and in $R_e-$scaled units. In the more massive galaxies, we can
reach larger physical radii, but the galaxies are also much
larger. Thus, in $R_e$ units we reach to larger distances in
the lower-mass galaxies.

From inspecting Figure \ref{fig:radial}, we choose to take a
measurement at $0.8-1 R_e$ that includes most (74) of the galaxies. We
also examine trends measured at $1.5 R_e$ for 58 galaxies, but we
systematically exclude the most massive galaxies in this sample due to
the very large size of these galaxies and the limited field-of-view of
the instrument. The trends with galaxy mass and size are somewhat
different when we consider fixed physical apertures. We determine that
we can reach $\sim 10$ kpc for 73 galaxies, with no real trend in mass
or \sig\ for the galaxies that are excluded. These do tend to be the
smallest galaxies.  We reach smaller physical radius in physically
smaller galaxies because we require S/N$>20$ to make the stellar
population measurements, and the smallest galaxies have very low
surface brightness at 10 kpc. We also take a measurement at 15 kpc for
45 galaxies, although in this case we systematically exclude the
lower-mass galaxies, which are both smaller on average and fainter,
thus making measurements at large physical radius most challenging.

To determine the measurements at each of these radii,
we simply take the measured radial profiles in \feh, and \afe, and
interpolate to the radius of choice by weighting each measurement by
radial distance (see Appendix A). These measurements at fixed radii
are the primary way that we investigate the spatial variation in
stellar populations, in conjunction with the radial gradients that we
describe in the next section.

\begin{figure*}
\begin{center}
\vskip -10mm
\vbox{
\includegraphics[width=2.5\columnwidth,angle=-90]{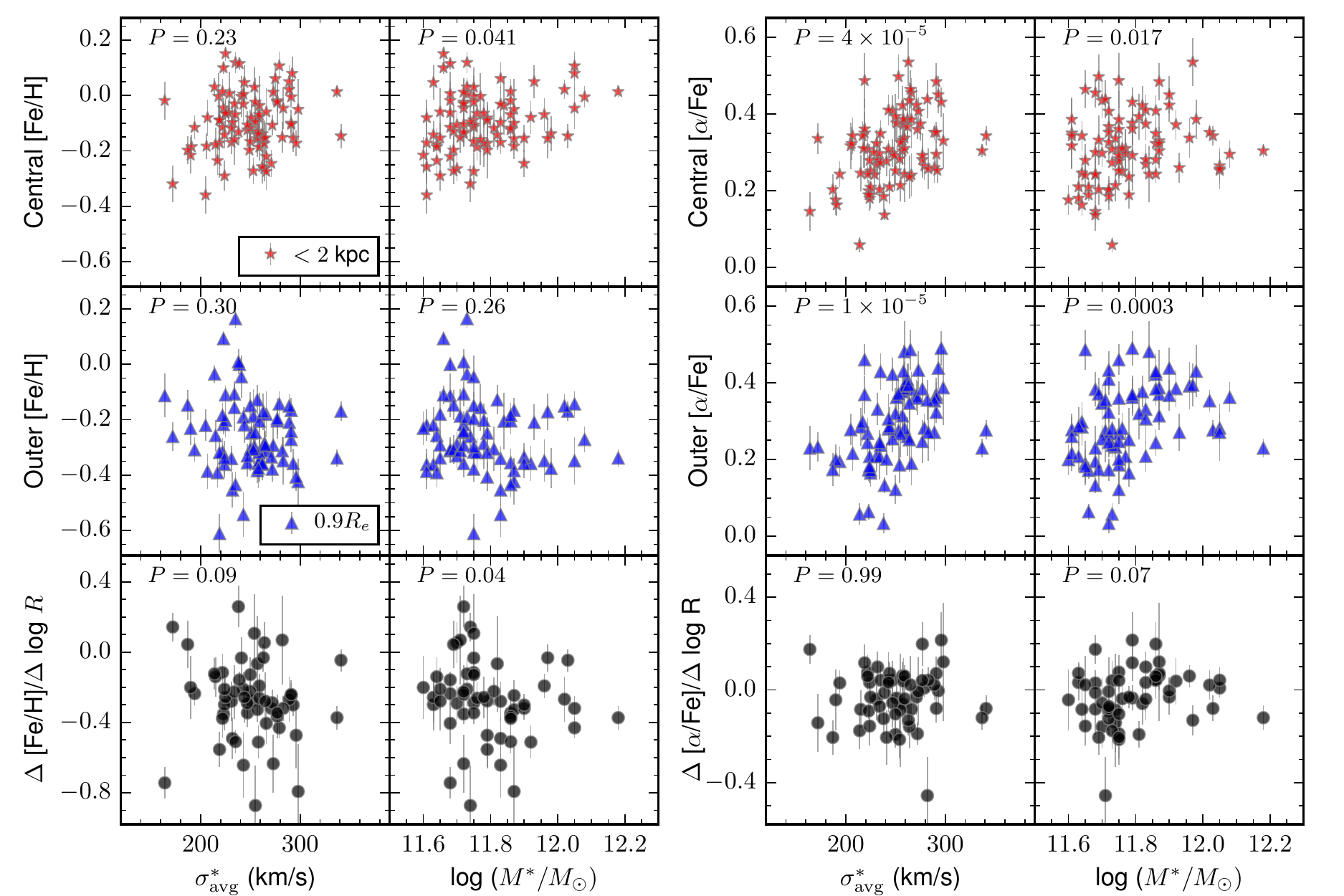}
}
\end{center}
\caption{
Correlations between stellar population parameters \feh\ (first 
two columns) and \afe\ (second two columns) with 
structural parameters \sig\ and \mstar. We show the correlations with 
the central values within $2$~kpc (top row; red stars), ``outer'' 
values at $0.9~R_e$ 
(second row; blue triangles) and finally the radial 
gradients (third row; grey circles). In all cases, the probability of 
the Spearman rank coefficient is shown in the top left; we take $P<0.05$ 
as significant (see also Table 2).}
\label{fig:afe-sig}
\end{figure*}

\subsection{Gradients in Stellar Populations}
\label{sec:stellarpops-gradients}

\begin{figure*}
\begin{center}
\vbox{
\includegraphics[width=0.7\columnwidth]{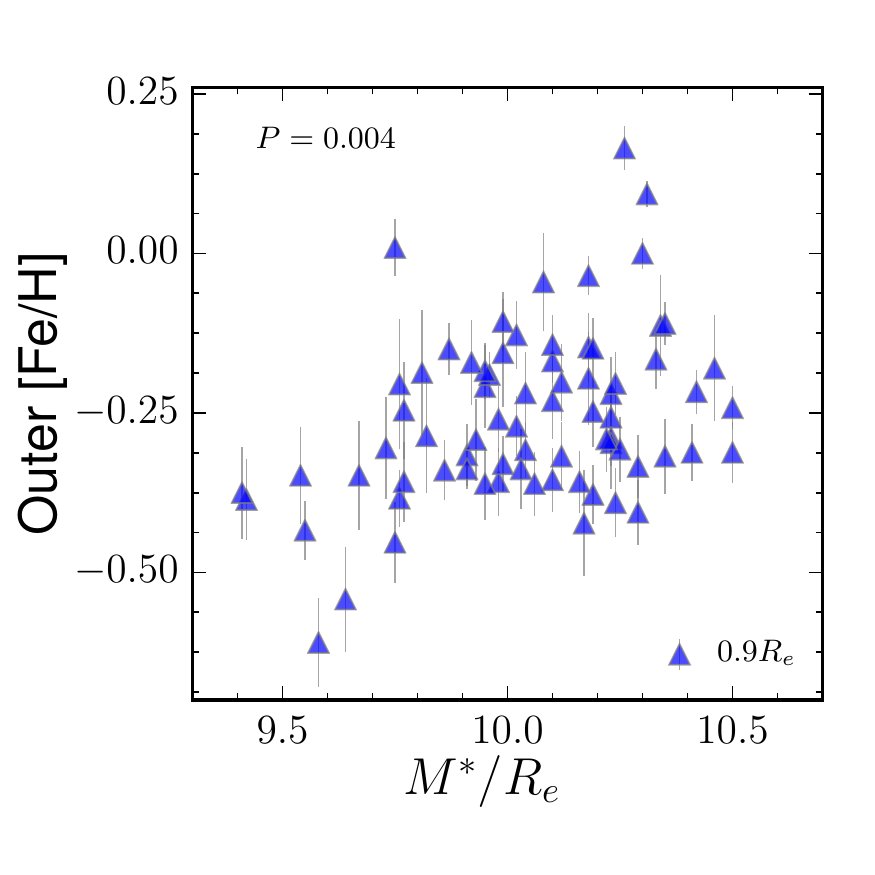}
\hskip -2mm
\includegraphics[width=0.7\columnwidth]{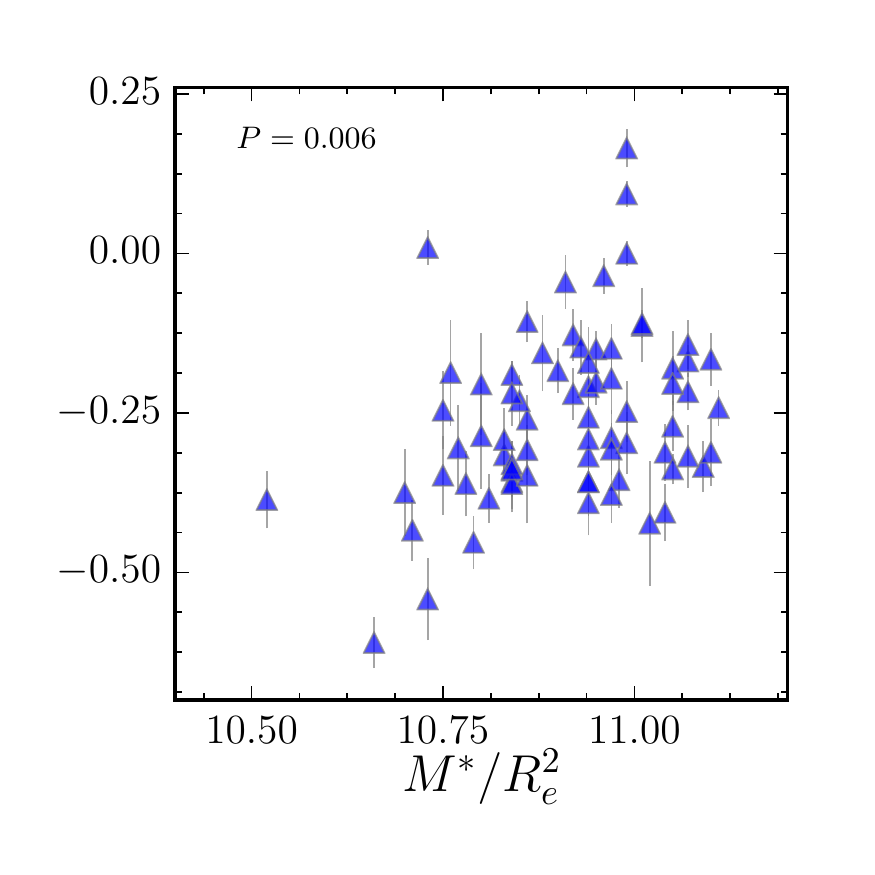}
\hskip -2mm
\includegraphics[width=0.7\columnwidth]{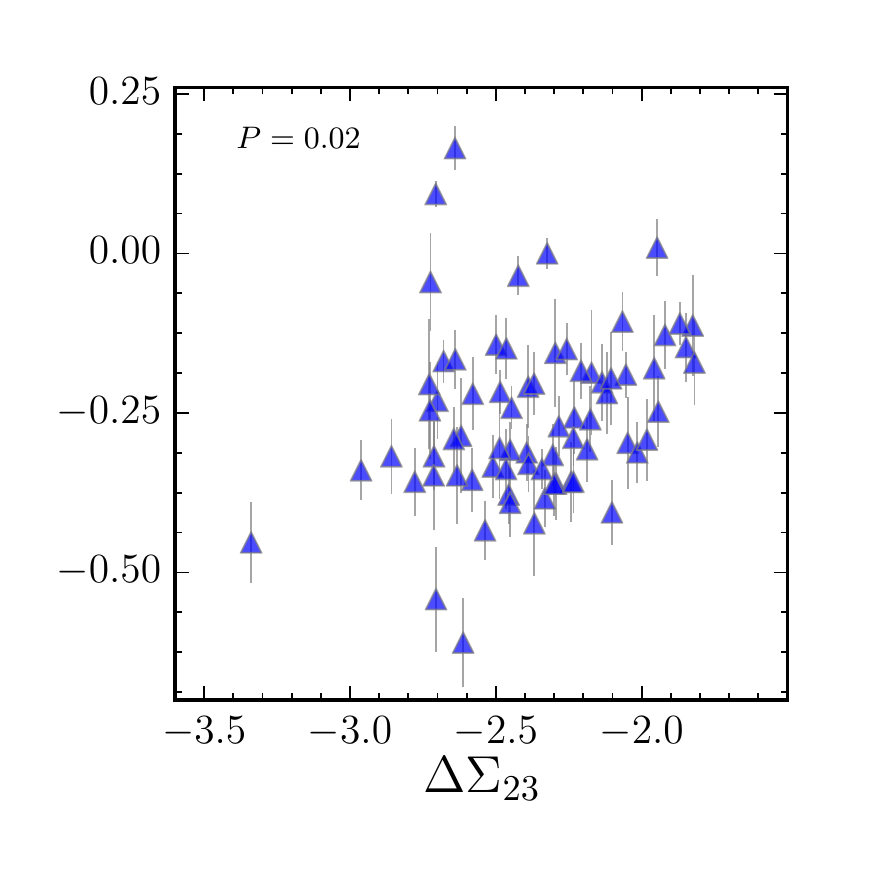}
}
\end{center}
\caption{The correlation between \feh\ measured at $0.9~R_e$ and 
a proxy for the gravitational potential ($M^*/R_e$; left), 
a proxy for the stellar surface density ($M^*/R_e^2$; middle), and the 
gradient in the surface brightness (right). Like 
\citet{baroneetal2018} we see a correlation with \feh\ at $0.9~R_e$, but we 
see no correlation with these parameters and \afe, nor with these parameters 
and \feh\ in the center of the galaxy. We also see a correlation with \feh\ 
at $0.9~R_e$ and $\Delta \Sigma_{23}$.
}
\label{fig:phidenscorr}
\end{figure*}

\begin{deluxetable*}{lllllllllll}
\tablenum{1}
\tablecolumns{11} 
\tabletypesize{\scriptsize}
\tablewidth{0pc}
\tablecaption{Galaxy Gradients \label{tablecorr}}
\tablehead{ 
\colhead{Gal} & \colhead{$M^*$} & \colhead{$\sigma$} & \colhead{$R_e$} & \colhead{$\Delta \Sigma_{23}$} & \colhead{[Fe/H]$_{\rm c}$} & \colhead{[Fe/H]$_{\rm o}$} & \colhead{$\Delta$[Fe/H]/$\Delta$ log $R$} & \colhead{[$\alpha$/Fe]$_{\rm c}$} & \colhead{[$\alpha$/Fe]$_{\rm o}$} & \colhead{$\Delta$[$\alpha$/Fe]/$\Delta$ log $R$}  \\
\colhead{(1)} & \colhead{(2)} & \colhead{(3)} & 
\colhead{(4)} & \colhead{(5)} & \colhead{(6)} & \colhead{(7)}
& \colhead{(8)} & \colhead{(9)} & \colhead{(10)} & \colhead{(11)}} 
\startdata
  NGC0057  &  11.8 &  251 &    6.3 & $-1.93\pm$0.10 &  $-0.07\pm$0.06 &  $-0.24\pm$0.06 &  $-0.27\pm$0.05 &   $0.38\pm$0.05 &   $0.36\pm$0.05 & $-0.034\pm$0.05 \\
  NGC0080  &  11.8 &  222 &    8.4 & $-2.30\pm$0.18 &  $-0.10\pm$0.05 &  $-0.31\pm$0.05 &  $-0.11\pm$0.11 &   $0.31\pm$0.03 &   $0.27\pm$0.04 &  $0.077\pm$0.07 \\
  NGC0315  &  12.0 &  341 &    9.2 & $-2.67\pm$0.08 &  $-0.14\pm$0.04 &  $-0.16\pm$0.03 &  $-0.04\pm$0.06 &   $0.34\pm$0.03 &   $0.28\pm$0.03 & $-0.079\pm$0.06 \\
  NGC0383  &  11.8 &  257 &    8.0 & $-1.91\pm$0.05 &  $-0.13\pm$0.05 &  $-0.12\pm$0.05 &  $-0.06\pm$0.27 &   $0.42\pm$0.05 &   $0.38\pm$0.04 & $-0.058\pm$0.09 \\
\enddata
\tablecomments{
The following is provided for guidance on form and content. The table is 
available in full online.
Col. (1): Galaxy name.
Col. (2): Stellar mass (\msun).
Col. (3): Average stellar velocity dispersion (\kms).
Col. (4): Effective (half$-$light) radius (kpc).
Col. (5): [Fe/H] value measured within $<2$~kpc (dex).
Col. (6): [Fe/H] value measured at $0.9~R_e$ (dex).
Col. (7): The gradient in the surface brightness per decade in radius, 
$\Delta \Sigma_{23}$, measured from $2-3 R_e$. 
Col. (5): [$\alpha$/Fe] value measured within $<2$~kpc (dex).
Col. (6): [$\alpha$/Fe] value measured at $0.9~R_e$ (dex).
Col. (6): The gradient in the \feh\ value, per decade in radius 
$\Delta$[Fe/H]/$\Delta$log~R.
Col. (6): The gradient in the \afe\ value, per decade in radius 
$\Delta$[$\alpha$/Fe]/$\Delta$log~R.
}
\end{deluxetable*}

We measure the radial gradients as linear fits to each stellar
population parameter as a function of the logarithm (base 10) of the
radius. The fits are linear and include the error bars in
the parameters. We show the radial profiles of all of the
  galaxies in Appendix A, with the gradient fits superposed.  As
emphasized above, the aperture used to measure the gradients can be
very important, so we experiment with measuring the gradients over
different radial ranges. We try adopting different inner and outer
radial coverage by measuring gradients between 1 and 10 kpc, 2 and 10
kpc, and then just truncating the inner radii at 1 or 2 kpc but
allowing the outer coverage to extend as far as possible.  We find
that there is a large variation in measured slopes when the inner
radius is changed. For \afe, taking the same outer radius but changing
the inner radius leads to a scatter in slopes of roughly 0.1. For the
\feh\ measurements, the scatter is larger (0.15 in slope). However, if
the inner radius is kept constant while the outer radius is varied,
the scatter is reduced by a factor of two. Thus, we choose to maximize
the radial coverage available for individual objects, and take a
standard inner radius of 1 kpc, but allow the outer radius to float
depending on the object. When we investigate trends with gradients, we exclude 
galaxies that have fewer than 10 radial points to define the 
gradient. 

The \feh\ gradients are generally negative, with a median value of
$-0.26$ dex per decade in radius (Figure \ref{fig:afe-sig}).  Only one
galaxy (NGC 6223) has a gradient that is consistent (at the $2~\sigma$
level) with having a positive gradient ($\Delta$[Fe/H]/$\Delta$ log
R$=0.26 \pm 0.12$). Unlike the \feh\ gradients, the \afe\ gradients
are nearly all consistent with being flat (Figure \ref{fig:afe-sig}),
with a median value ($-0.03$) that is only very slightly
negative. There are four galaxies that have \afe\ gradients that are
$>3~\sigma$ away from being zero.  One (NGC 2256) has a positive
gradient ($\Delta$[$\alpha$/Fe]/$\Delta$ log R$=0.23 \pm 0.07$). Three
have negative slopes (NGC 1272, UGC 02783, NGC 1453) with
$\Delta$[$\alpha$/Fe]/$\Delta$ log R$=-0.19 \pm 0.06, -0.23 \pm 0.08,
-0.19 \pm 0.05$. These outliers are likely to be real; if we assume
that all \afe\ gradients have the same intrinsic value of $-0.03$, and
then perturb each measurement by its uncertainty, we expect to detect
three $>3 \sigma$ outliers in the full sample only 1\% of the time,
and we find four outliers. On the other hand, the one positive
\feh\ gradient could be marginally consistent with pure scatter. If we
assume that all the \feh\ gradients are intrinsically equal to the
mean $-0.3$ dex per decade, then given our error bars we expect to
find a positive gradient 2\% of the time.

\begin{deluxetable}{lcccll}
\tablenum{2}
\tablecolumns{6} 
\tabletypesize{\scriptsize}
\tablewidth{0pc}
\tablecaption{Structural Correlations \label{tablecorr}}
\tablehead{ 
\colhead{Stellar Pop} & \colhead{Radius} & \colhead{Gal. Prop.} & 
\colhead{$N_{g}$} & \colhead{$\rho$} & \colhead{$P$} \\
\colhead{(1)} & \colhead{(2)} & \colhead{(3)} & 
\colhead{(4)} & \colhead{(5)} & \colhead{(6)}} 
\startdata
   \afe\ &  $<2$ &  $\sigma$ & 87 &  0.4 &  $4 \times 10^{-5}$ \\
   \afe\ &  $0.9 R_e$ & $\sigma$ & 74 &  0.5 & $1 \times 10^{-5}$ \\
   \afe\ &  9kpc &  $\sigma$ & 73 & 0.4 & 0.00018 \\
   \afe\ &  1.5$R_e$ &  $\sigma$ & 58 & 0.6  & $4 \times 10^{-6}$ \\
   \afe\ &  15kpc &  $\sigma$ & 45 & 0.4 & 0.003 \\
   \afe\ &  $<2$ & $M^*$ & 87 & 0.3 &  0.02 \\
   \afe\ &  $0.9 R_e$ & $M^*$ & 74  & 0.4 & 0.0003 \\
   \afe\ &  9kpc & $M^*$ &  73 & 0.4 & 0.0003 \\
   \afe\ &  15kpc & $M^*$ &  45 & 0.4 & 0.02 \\
   \afe\ &   $<2$ kpc &  $h4$ & 87 & 0.4 &  0.0002 \\
   \afe\ &  $0.9 R_e$ &  $h4$ & 74 &  0.5 &  $2 \times 10^{-5}$ \\
   \afe\ &  9~kpc &  $h4$ & 73 & 0.4 &  0.0001 \\
   \afe\ &  $1.5 R_e$ &  $h4$ & 74 &  0.5 &  $7 \times 10^{-5}$ \\
   \afe\ &  15~kpc &  $h4$ & 73 &  0.4 &  0.003 \\
\hline
   \feh\ &  $<2$ kpc  & $M^*$ & 87 &  0.2 & 0.04 \\
   \feh\ &  $0.9 R_e$ &   $M^*/R$ & 74  & 0.3 & 0.006 \\
   \feh\ &  $0.9 R_e$ & $M^*/R^2$ & 74  & 0.3 & 0.004 \\
   \feh\ &  $0.9 R_e$ &  $h4$ & 73 & $-0.3$ & 0.003 \\
  \feh\ &  $0.9 R_e$ &    $\Delta \Sigma_{23}$ & 81 & 0.3 & 0.02 \\
\enddata
\tablecomments{
We investigate correlation coefficients (using Spearman's $\rho$) 
between \feh\ and \afe, and their gradients, 
and the structural parameters $\sigma$, outer gradient in the 
$\sigma$ profile ($\gamma_o$),
$M^*$, $M^*/R$, $M^*/R^2$, and the slope of the surface-brightness 
profile. Correlation coefficients with $P<0.05$ are included here. 
Col. (1): Stellar population parameter.
Col. (2): Radial extraction radius.
Col. (3): Galaxy property.
Col. (4): Number of galaxies included in the correlation test.
Col. (5): Spearman $\rho$.
Col. (6): Spearman probability.
}
\end{deluxetable}

\section{Scaling Relations Between Stellar Populations and Galaxy Properties}
\label{sec:structure}

\begin{figure*}
\begin{center}
\vbox{
\includegraphics[width=0.75\columnwidth]{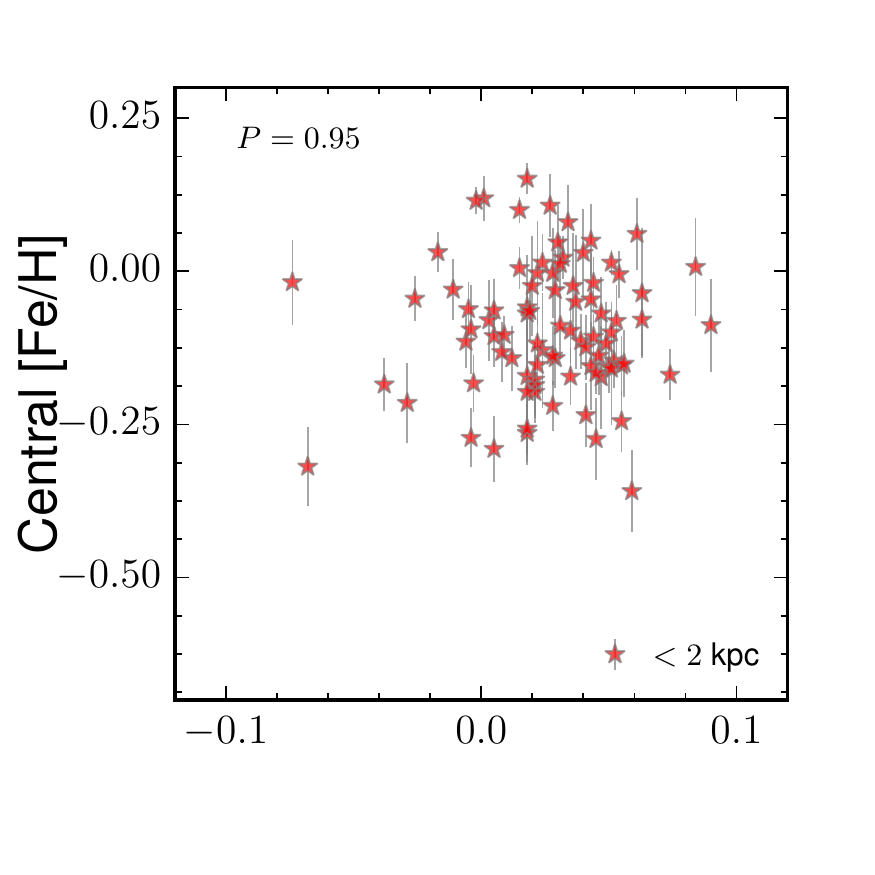}
\hskip -2mm
\includegraphics[width=0.75\columnwidth]{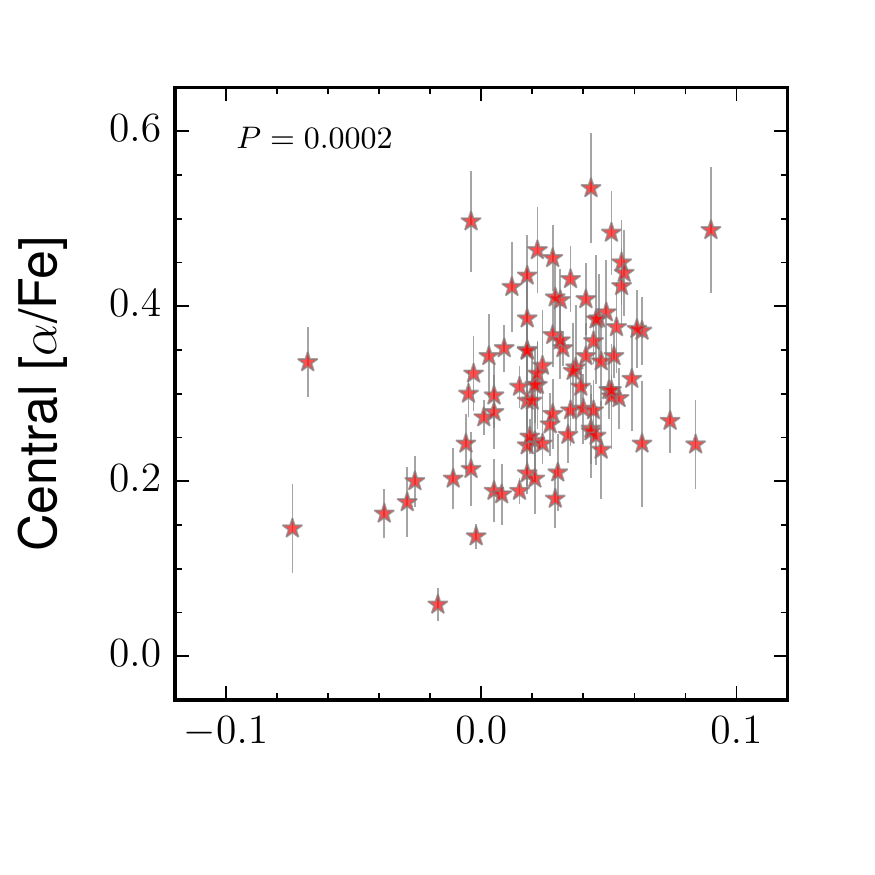}
}
\vskip -9.5mm
\vbox{
\includegraphics[width=0.75\columnwidth]{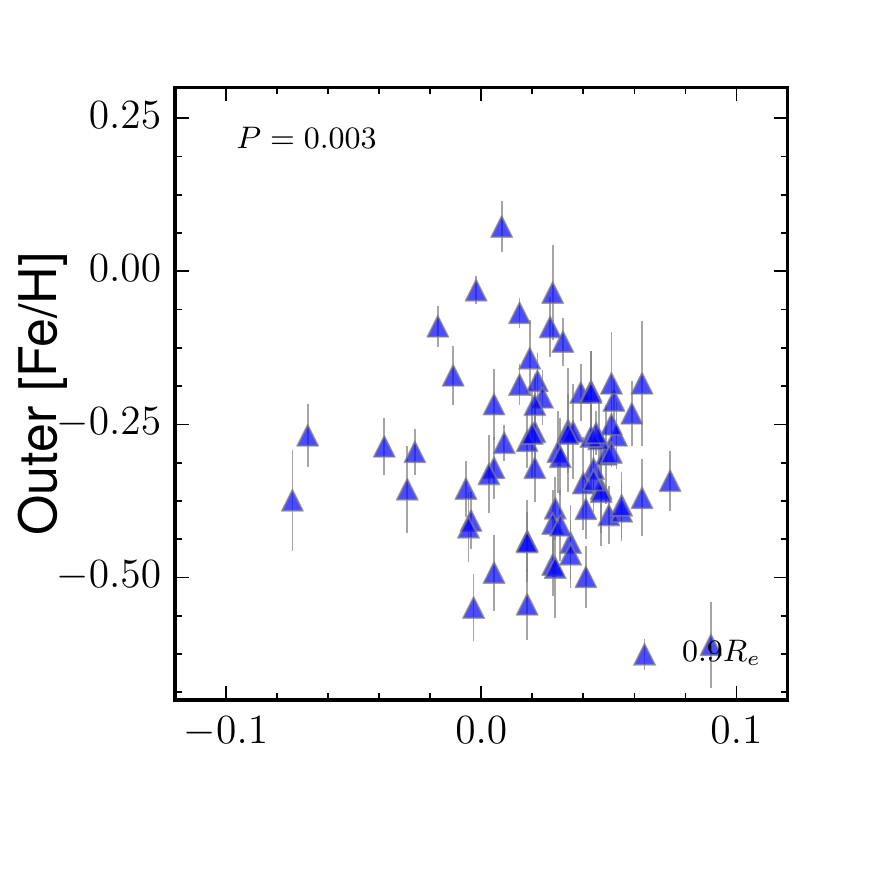}
\hskip -1mm
\includegraphics[width=0.75\columnwidth]{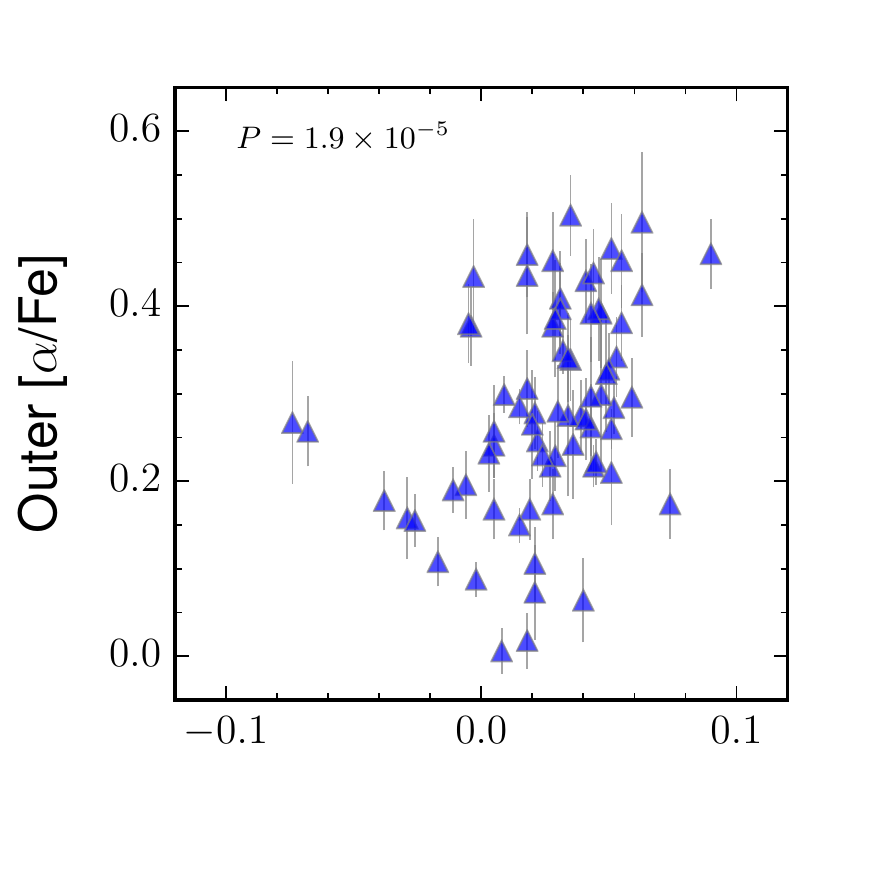}
}
\vskip -9.5mm
\vbox{
\includegraphics[width=0.75\columnwidth]{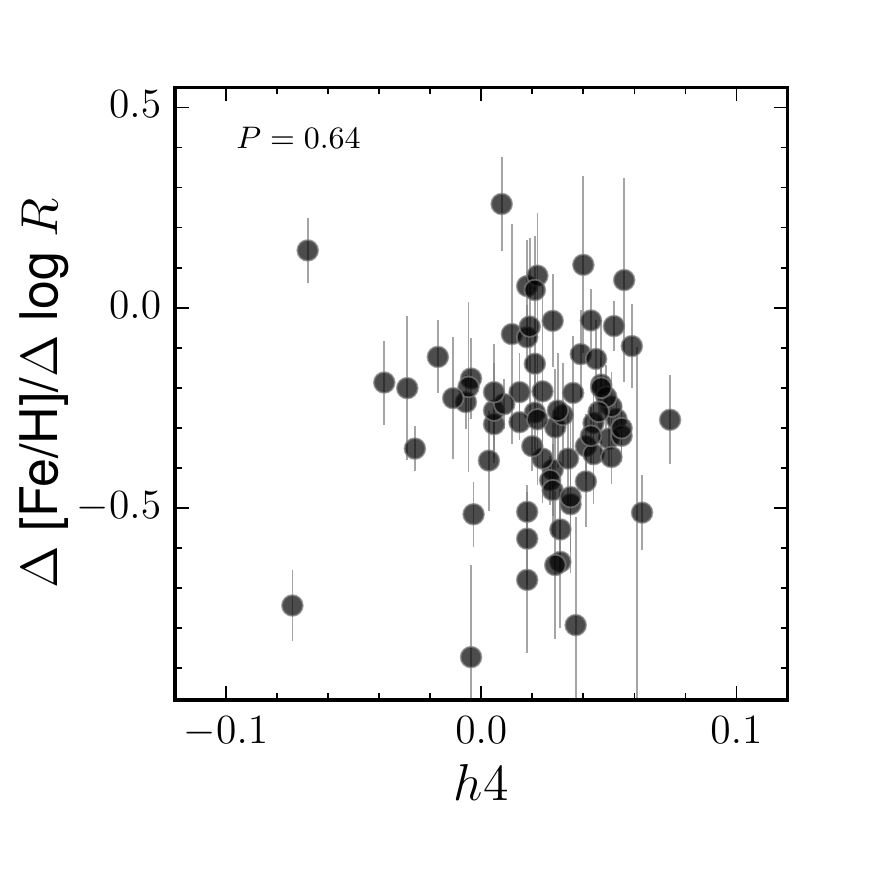}
\hskip -1mm
\includegraphics[width=0.75\columnwidth]{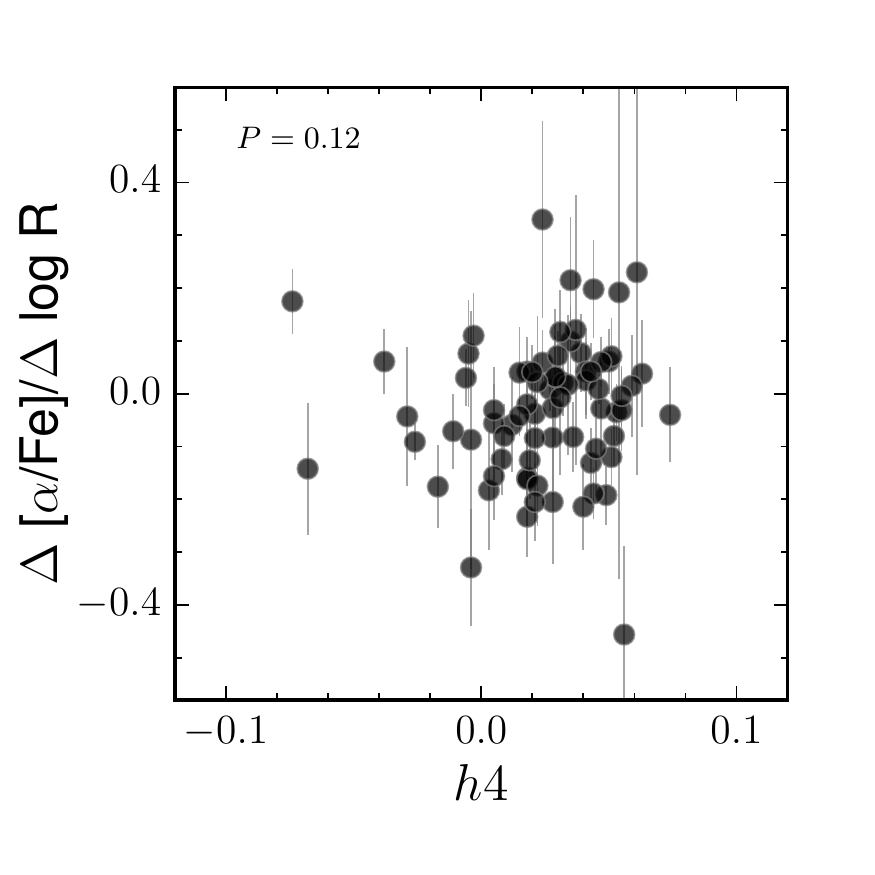}
}
\end{center}
\caption{{\it Left}: The relationship between 
the higher-order moment of the LOSVD, $h4$, and 
\feh\ as measured in the galaxy center ($<2$~kpc), 
$0.9~R_e$, and the gradient 
in \feh. We see no correlation with the central values but a clear and strong 
correlation with \feh\ as measured at $R_e$. This correlation is 
not seen with \sig, so appears to be an independent relationship. 
{\it Right}: As at left, but now the relationship between $h4$ and \afe\ as 
measured with the same apertures. The correlations are comparably strong 
as those seen 
with \sig, and in the same sense (higher $h4$ corresponds to higher \afe).  
}
\label{fig:h4corr}
\end{figure*}

Early-type galaxies are known to exhibit scaling relationships between
their structural properties and their stellar populations
\citep[e.g.,][]{trageretal2000b,thomasetal2005,gravesetal2009}. Only
recently, with the advent of large IFS surveys, has it become possible
to examine these scaling relationships outside of the central regions
of the galaxies
\citep[e.g.,][]{greeneetal2015,vandesandeetal2018,baroneetal2018}, and
since there are gradients in metallicity with radius, different trends 
have different dependence on the aperture used.

In this section, we focus on possible trends between structural
properties of our galaxies and their stellar populations in the center
and at large radius. Specifically, we examine stellar mass as inferred
from the $K-$band magnitude, using a dynamically derived mass-to-light
ratio \citep[see details in ][]{maetal2014,cappellari2013}. We also
look at the stellar velocity dispersion, and since we are interested
in our galaxies on large scales, we use the luminosity weighted average 
\sig\ over all Mitchell bins within $R_e$ 
\citep[Table 1, col 12 in ][]{vealeetal2017a}. We also
consider proxies for the gravitational potential $\Phi \propto M^*/R_e$
and the stellar surface density $\Sigma \propto M^*/R_e^2$ following
\citet{baroneetal2018}, who study a galaxy sample of (typically) lower
$M^*$ with SAMI. Motivated by the Illustris works of
\citet{pillepichetal2014} and \citet{cooketal2016}, we investigate
correlations between the stellar population parameters and the slope
of the surface brightness profile 
$\Delta \Sigma_{23} \equiv \Delta \Sigma/\Delta \rm{log} R (2-3 R_e)$ 
(\S \ref{sec:sigma}; Figure \ref{fig:phidenscorr}).

\subsection{Structural Correlations with the Galaxy Centers}

By virtue of our IFS data, we have stellar population measurements out
to $> 10$ kpc ($\sim R_e$ or beyond; Fig.\ 1) for the majority of
galaxies in our sample.  Our primary interest is to investigate the
stellar content in the outer parts of the galaxies, but first we show
briefly that our central measurements are consistent with prior work
\citep[for more detailed comparisons, see][]{greeneetal2012,
  greeneetal2013}. We construct ``central'' stellar population
measurements using all of the fibers contained within 2 kpc of the
galaxy center.  These are luminosity-weighted measurements, to mimic
the process for individual SDSS fibers. We then ask
whether our central measurements correlate with \sig\ or \mstar.  To
test for correlation, we use the non-parametric Spearman correlation
coefficient.  The results of our correlation tests are presented in
Table 2 and Figure \ref{fig:afe-sig}, where we only include 
``significant'' correlations with a probability $P<0.05$ of the null 
hypothesis. 

In keeping with prior results, we find very strong correlations
between \afe\ and \sig\ ($\rho = 0.5$, $P=1 \times 10^5$ of a null result),
significant correlations between \afe\ and \mstar\ ($\rho = 0.26$,
$P=0.017$), and weak to no correlation between \feh\ and
\mstar\ ($\rho = 0.22$, $P=0.04$) or \sig\ 
\citep[$\rho = 0.13$, $P=0.23$; e.g.,][]{trageretal2000a,gravesetal2009,wakeetal2012,
  conroyetal2014}. 

We also find an interesting hint that \afe\ saturates above
\mstar$=10^{11.8}$~\msun. At the lowest masses covered in our sample,
\afe\ shows an increasing trend, but then seems to flatten out at the
highest masses. Quantitatively, galaxies in a stellar mass bin of
\mstar$=10^{11.8} - 10^{12}$~\msun, and all galaxies more massive than
this limit, both have a consistent weighted mean \afe$=0.32 \pm
0.01$. This consistency at high mass holds independently of exactly
how we divide the two mass bins. This convergence may be a sign that
major mergers are needed to make these most massive galaxies.
Simulations also show that at the highest stellar masses, the
predominance of more major mergers leads to a convergence of
properties for high-mass central and satellite galaxies
\citep[e.g.,][]{wetzeletal2013}.

By virtue of our IFU data and high signal-to-noise ratio (SNR) spectra, 
we are able to look for correlations with other kinematic tracers 
beyond \sig. We see intriguing
correlations with the higher-order moments of the line-of-sight
velocity distribution (LOSVD).  Using a Gauss-Hermite decomposition,
we model the LOSVD with velocity, velocity dispersion, and four
higer-order moments ($h3-h6$) as described in detail in
\citet{vealeetal2017a}. Specifically, for spectrum $f(v)$, mean velocity $V$, 
velocity dispersion $\sigma$, and the number 
of higher orders $n=6$, we have:
$$f(v) \propto \frac{\rm{exp}\left[\frac{(v-V)^2}{\sigma^2}\right]}{\sqrt{2 \pi \sigma^2}}
\left[1 + \Sigma_{m=3}^{n} h_m H_m \left(\frac{v-V}{\sigma}\right) \right]  $$

And the $H_m(x)$ are the Hermite polynomials:

$$ H_m(x) = \frac{1}{\sqrt{m!}} \rm{exp}[x^2] \left(-\frac{1}{\sqrt{2}} \frac{\partial}{\partial x} \right)^m \rm{exp}[-x^2] $$

The third term $h3$ is the skewness, and the fourth term $h4$ is the 
kurtosis. To maximize SNR, we use 
a light-weighted average $h4$ measurement 
\citep[see details in ][]{vealeetal2017a}. In the galaxy 
center, we see a correlation between $h4$ and \afe\ that is as 
strong ($\rho = 0.5$, $P=2 \times 10^5$) as the correlation 
between \sig\ and \afe\ (Figure \ref{fig:h4corr}). We see no correlation between 
$h4$ and central values of \feh\ ($\rho = -0.0073$, $P=0.95$). 

\subsection{Structural Correlations with the Outskirts}

We then extend the above analysis to the outer regions of the galaxies
and a broader range of structural parameters.  In all cases, we seek
correlations between the stellar populations measured 
at fixed $R_e$-scaled radii ($0.9 R_e$ and $1.5 R_e$) and fixed
physical radii (9 kpc and 15 kpc), and gradients measured beyond 1 kpc (\S
\ref{sec:stellarpops-gradients}). All significant correlations are 
included in Table 2.
The correlations we saw between central \afe\ and both \sig\ and
\mstar\ persist when we investigate the outer parts of the
galaxies (\sig-\afe\ at $0.9 R_e$: $\rho = 0.5$, $P=10^{-5}$; 
\mstar-\afe\ at $0.9 R_e$: $\rho = 0.4$, $P=0.0003$). 
This is not surprising, since we do not measure significant gradients in \afe\ (Figure
\ref{fig:afe-sig}). In contrast, no compelling correlations between
\sig\ or \mstar\ emerge with \feh\ or its gradients at large
radius (\sig-\feh\ at $0.9 R_e$: $\rho = -0.12$, $P=0.3$; 
\mstar-\feh\ at $0.9 R_e$: $\rho = -0.13$, $P=0.26$). We will discuss the 
correlations with higher-order moments in the following section.

Like \citet{baroneetal2018}, we find a significant correlation between
$M^*/R_e$ ($\rho=0.3,P=0.006$; a proxy for the gravitational
potential) and $M^*/R_e^2$ ($\rho=0.3,P=0.004$; the stellar surface
density) with \feh\ measured at $0.9 R_e$ (Figure
\ref{fig:phidenscorr}; Table 2), and in common with that work, we find
no such correlation for the central stellar populations, only those
measured at larger radius. Higher metallicity in higher surface mass
density and/or higher gravitational potential strongly suggests that
the stars in question were formed in a similar potential to their
current one, and that the metal retention rate is set by the
gravitational potential.

We next examine the correlation between stellar populations and
$\Delta \Sigma_{23}$, the gradient in the surface brightness profile
measured between $2-3 R_e$. We are motivated by results from Illustris
\citep{vogelsbergeretal2013} showing that the surface brightness slope
grows shallower as galaxies accrete more of their stars from external
galaxies \citep{pillepichetal2014}. \citet{cooketal2016} expand on
this finding to show that the metallicity gradients also grow
shallower as the amount of accretion increases. We do find a very
similar range in surface brightness slopes in our data to that
presented by \citet{cooketal2016} for a similar stellar mass
range. However, we do not detect a significant correlation between the
gradients in stellar population parameters and the gradients in
surface brightness (with $P>0.3$ in both cases). Nor do we find a
correlation between $\Delta \Sigma_{23}$ and \afe\ measured at any
radius ($P>0.2$ in all cases). $\Delta \Sigma_{23}$ is postively
correlated with \feh\ measured at $0.9 R_e$ ($\rho=0.28, P=0.02$) and
\feh\ at $1.5 R_e$ ($\rho=0.26,P=0.05$).

Finally, we checked for correlations between the stellar
population gradients and the ratio of rotational to dispersion support
\citep[$\lambda$;][]{emsellemetal2007,vealeetal2017a} as well as
looking for any interesting stellar population properties for galaxies
with counter-rotating componenets \citep{eneetal2018}, but we did not
find any significant trends. 

\subsection{The correlation between $h4$ and outer stellar populations}

As described above, we are able to robustly measure not only $V$ and
\sig, but also deviations from Gaussian LOSVDs from our high SNR
data. We are particularly interested in the kurtosis, $h4$. High
kurtosis can be a sign of radial anisotropy, which in turn may point
to stellar accretion events
\citep[e.g.,][]{wuetal2014,amorisco2017}. We investigate possible
correlations between $h4$ and the stellar populations in the outer
parts of the galaxy (Figure \ref{fig:h4corr}). To maximize the SNR, we
use the light-weighted average $h4$ measurement from
\citet{vealeetal2018}.

We find that the global $h4$ measurement correlates with the $r=0.9
R_e$ measurements of both \feh\ ($\rho=-0.33, P=0.003$) and
\afe\ ($\rho=0.48$, $P=2 \times 10^{-5}$). The correlation between
\afe\ and $h4$ is comparably strong to that between \afe\ and \sig. In
the case of \feh, we do not see a correlation with \sig\ at all, but
detect a very significant correlation with $h4$ at $R \approx 0.9 R_e$. 
This correlation persists even if we remove the four points with
the lowest \feh$<-0.4$.  Thus, we do not believe that the correlation
with $h4$ is somehow derivative of the correlation with
\sig. Likewise, there is a known correlation between $h4$ and stellar
mass in this sample \citep{vealeetal2017a}, but \feh\ at $0.9 R_e$ and
\mstar\ are not correlated either ($\rho=-0.13,P=0.26$). To double check 
that the correlation between stellar populations and $h4$ is not derivative of 
correlations with stellar mass, we also perform a linear fit to \afe\ and \feh\ 
versus stellar mass, dividing into two $h4$ bins. We find significant differences 
in zero-point for both \afe\ ($0.22 \pm 0.006$ for $h4<0.029$; 
$0.29 \pm 0.007$ for $h4>0.029$) and \feh\ ($-0.15 \pm 0.007$ for low $h4$; 
$-0.28 \pm 0.009$ for high $h4$) at $0.9 R_e$. This difference confirms our claim 
that even at fixed mass, there are real differences in the stellar populations as a 
function of $h4$.

One concern is that template mismatch may cause an apparent
correlation between $h4$ and stellar population properties, where $h4$
rises to compensate, for instance, if the template library does not
have stars with high \afe\ and low \feh. To investigate this possible
bias, we look for correlation between the local values of $h4$, \feh,
and \afe\ measured in each spatial bin of individual galaxies.
Restricting our attention to systems with more than 15 radial bins, we
find that only $\sim 6\%$ show correlation between local $h4$ and
\afe\ (with the majority showing an anticorrelation, rather than the
positive correlation seen for the ensemble) and $\sim 14\%$ show
correlation between \feh\ and $h4$, split between positive and
negative correlation. We thus conclude that the correlations we see
between $h4$ and the outer stellar population properties are not
caused by template mismatch.

\begin{figure*}
\begin{center}
\vbox{
\includegraphics[width=0.75\columnwidth]{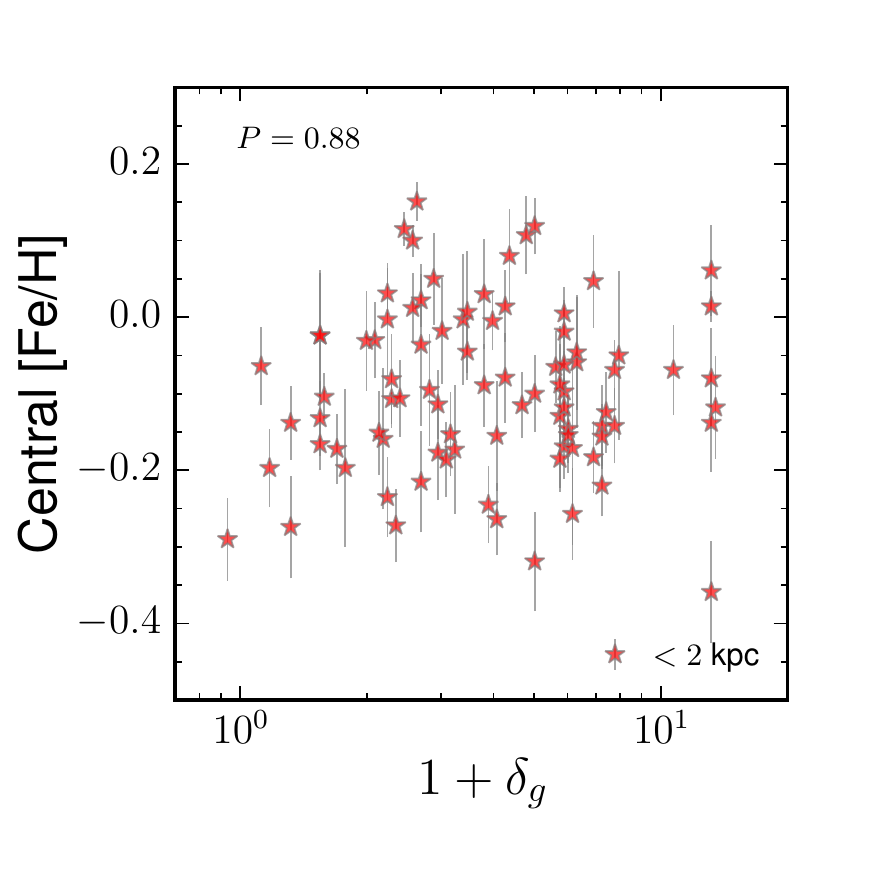}
\hskip -2mm
\includegraphics[width=0.75\columnwidth]{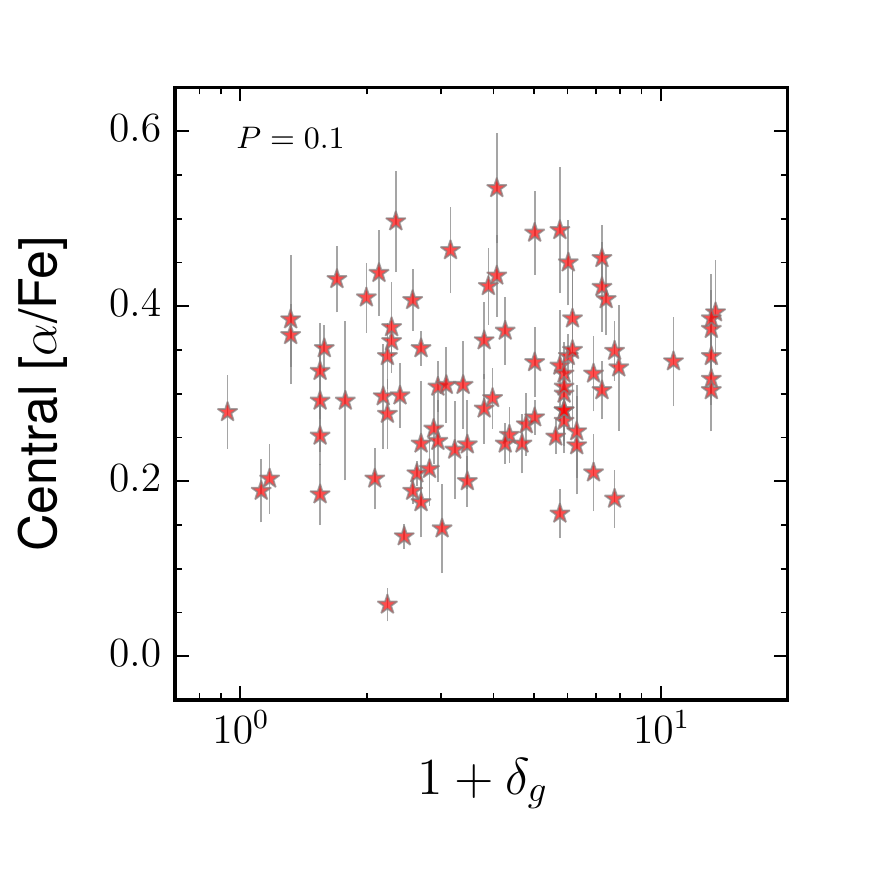}
}
\vskip -9.5mm
\vbox{
\includegraphics[width=0.75\columnwidth]{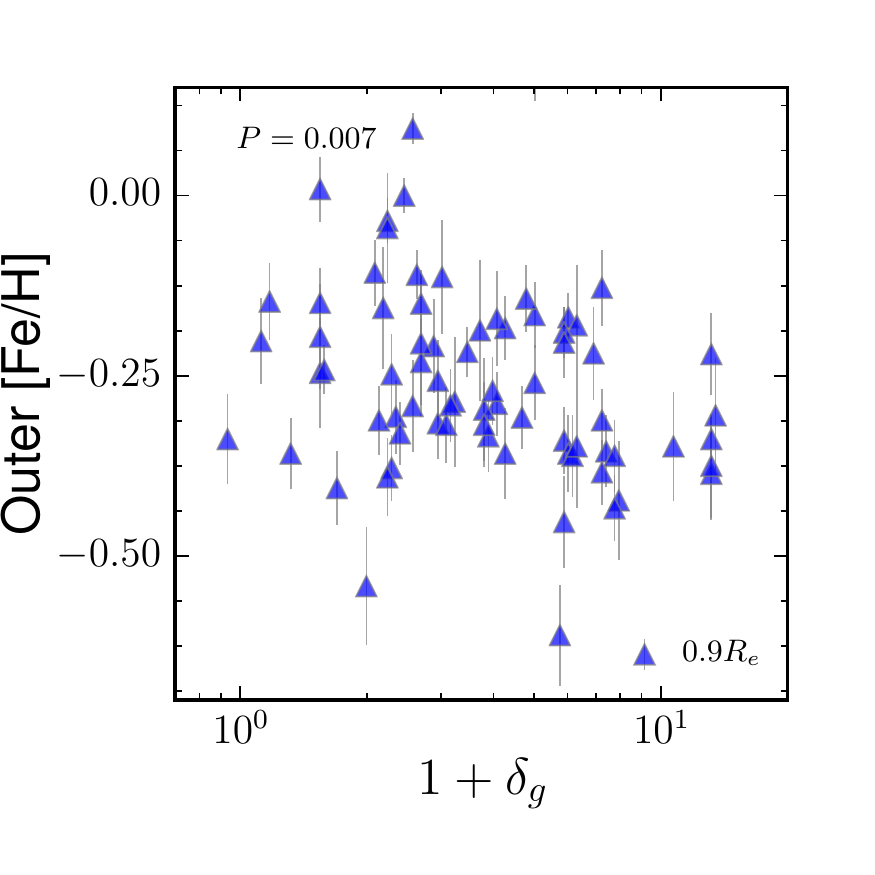}
\hskip -2mm
\includegraphics[width=0.75\columnwidth]{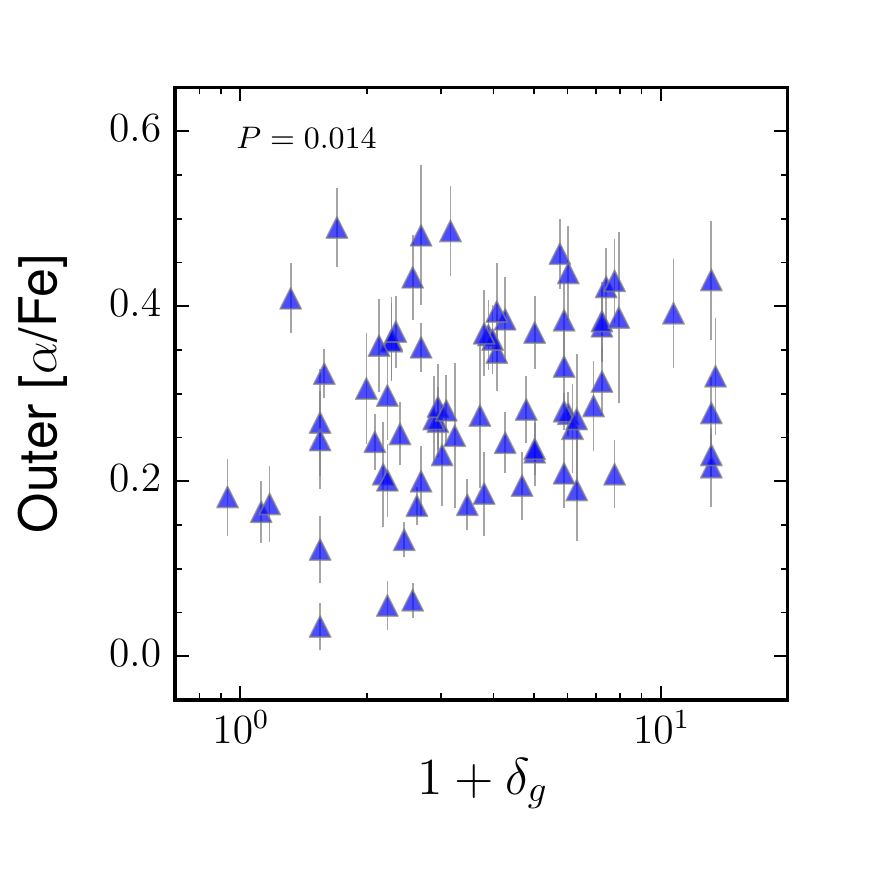}
}
\vskip -9.5mm
\vbox{
\includegraphics[width=0.75\columnwidth]{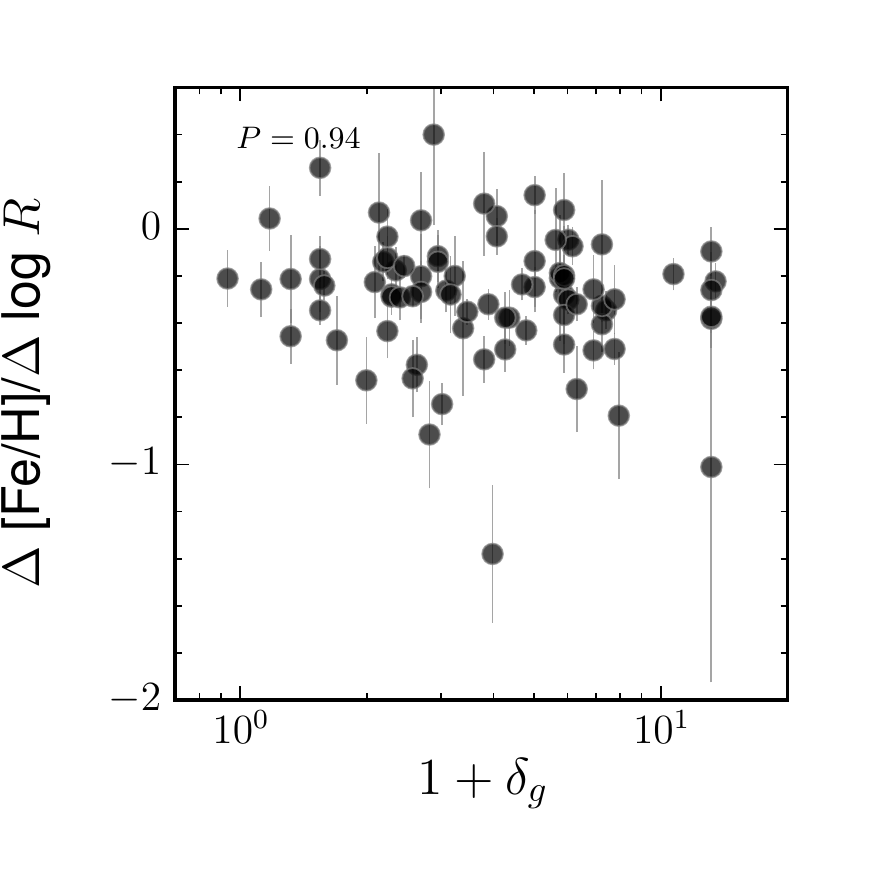}
\hskip -2mm
\includegraphics[width=0.75\columnwidth]{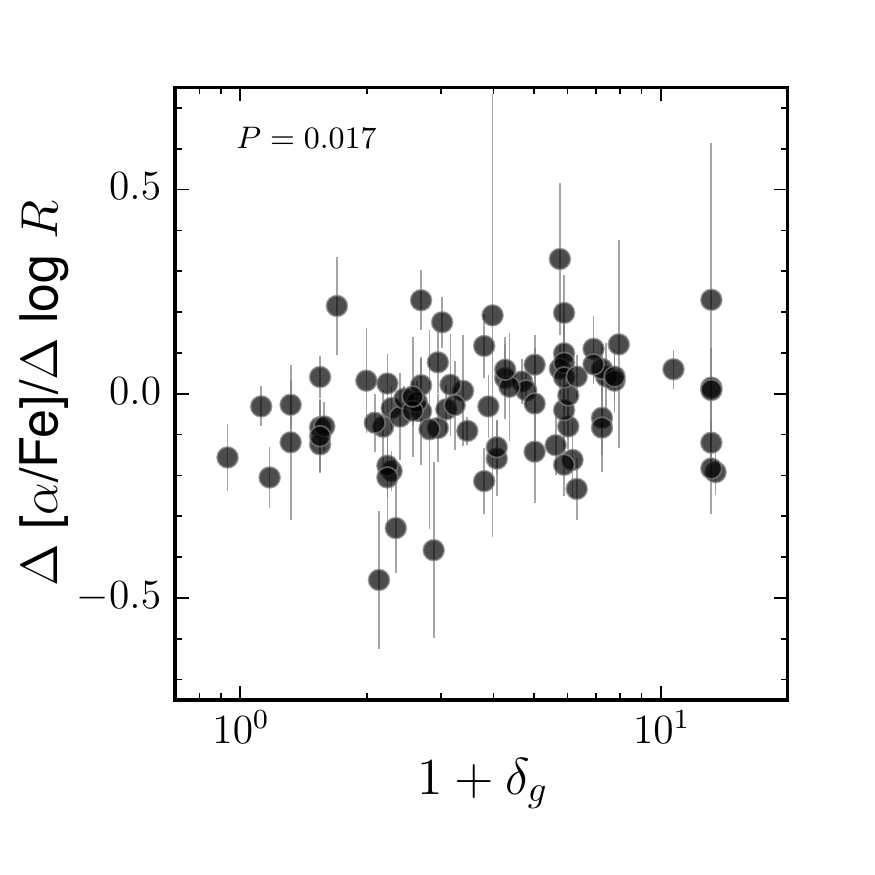}
}
\end{center}
\caption{Correlations between large-scale environment ($\delta_g$) and 
\feh\ (left) and \afe\ (right). We look in the central regions ($<2$~kpc; top), 
the outer region ($0.9~R_e$; middle), and at the gradients measured 
at radii $>1$ kpc (bottom). We 
measure clear correlations between both the abundances and the abundance ratios 
and $\delta_g$. We do not see a correlation between gradients in \feh\
and large-scale environment, but we do see a weak correlation between \afe\ gradients 
and $\delta_g$. 
}
\label{fig:env}
\end{figure*}

\begin{deluxetable}{lcccrr}
\tablenum{3}
\tablecolumns{6} 
\tabletypesize{\scriptsize}
\tablewidth{0pc}
\tablecaption{Environmental Correlations \label{tablecorr}}
\tablehead{ 
\colhead{Stellar Pop} & \colhead{Radius} & \colhead{Env.} & \colhead{$N_g$} & \colhead{$\rho$} & \colhead{$P$} \\
\colhead{(1)} & \colhead{(2)} & \colhead{(3)} & 
\colhead{(4)} & \colhead{(5)} & \colhead{(6)}} 
\startdata
   \afe\ &   $0.9Re$  &  $\delta_g$ & 74 &  0.3 & 0.014 \\
   \afe\ &  9kpc &  $\delta_g$ & 73 &  0.3 & 0.006 \\
   \afe\ &   $1.5Re$  &  $\delta_g$ & 58 &  0.6 & $5 \times 10^{-6}$ \\
   \afe\ &   $0.9Re$  &     $\nu$ & 74 &  0.3 & 0.015 \\
   \afe\ &  9kpc &     $\nu$ & 73 &  0.3 & 0.012 \\
   \afe\ &   $1.5Re$  &     $\nu$ & 58 &  0.4 & 0.0005 \\
 d\afe/dr & \nodata & $\delta_g$ & 77  & $0.3$ & 0.02 \\
\hline
   \feh\ &   $0.9Re$  &  $\delta_g$ & 73 & $-0.3$ & 0.007 \\
   \feh\ &  15kpc  &  $\delta_g$ & 45 &  $-0.3$ & 0.036 \\
   \feh\ &   $0.9Re$  &     $\nu$ & 73 &  $-0.3$ & 0.004 \\
\enddata
\tablecomments{
Correlations between the stellar population 
parameters and two proxies for the larger-scale galaxy density, 
$1+\delta_g$ and $\nu$.
Col. (1): Stellar population parameter.
Col. (2): Radial extraction radius.
Col. (3): Environmental measure.
Col. (4): Number of galaxies included in the calculation.
Col. (5): Spearman $\rho$.
Col. (6): Spearman probability.
}
\end{deluxetable}

\begin{figure*}
\begin{center}
\vbox{
\includegraphics[width=0.8\columnwidth]{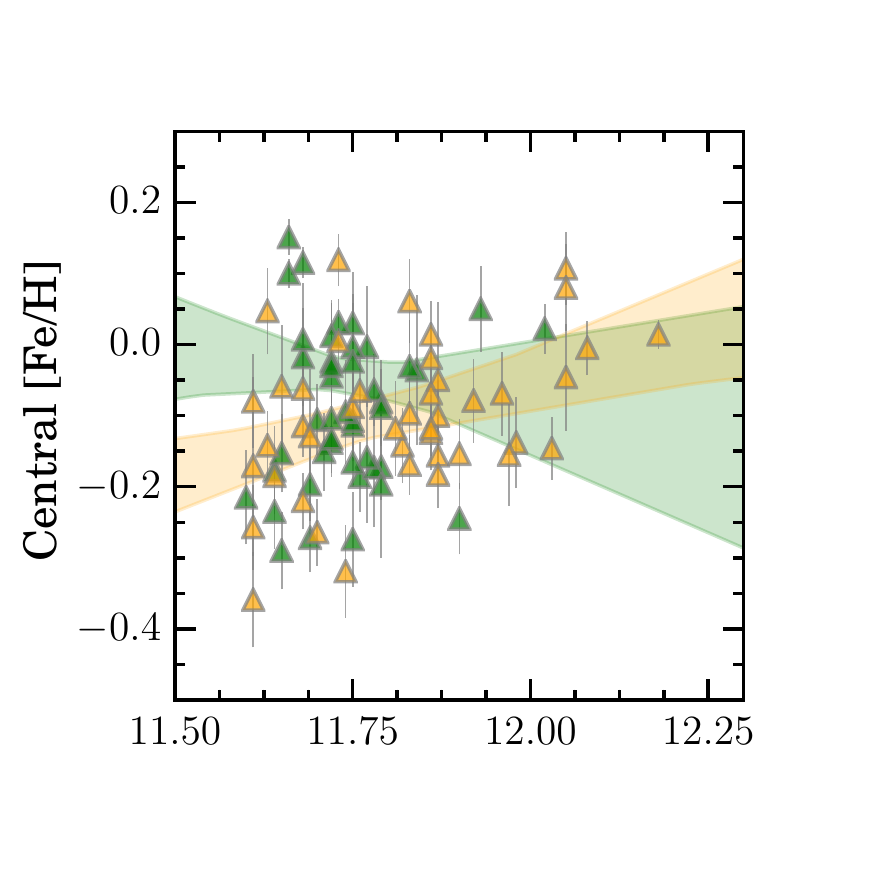}
\hskip -3mm
\includegraphics[width=0.8\columnwidth]{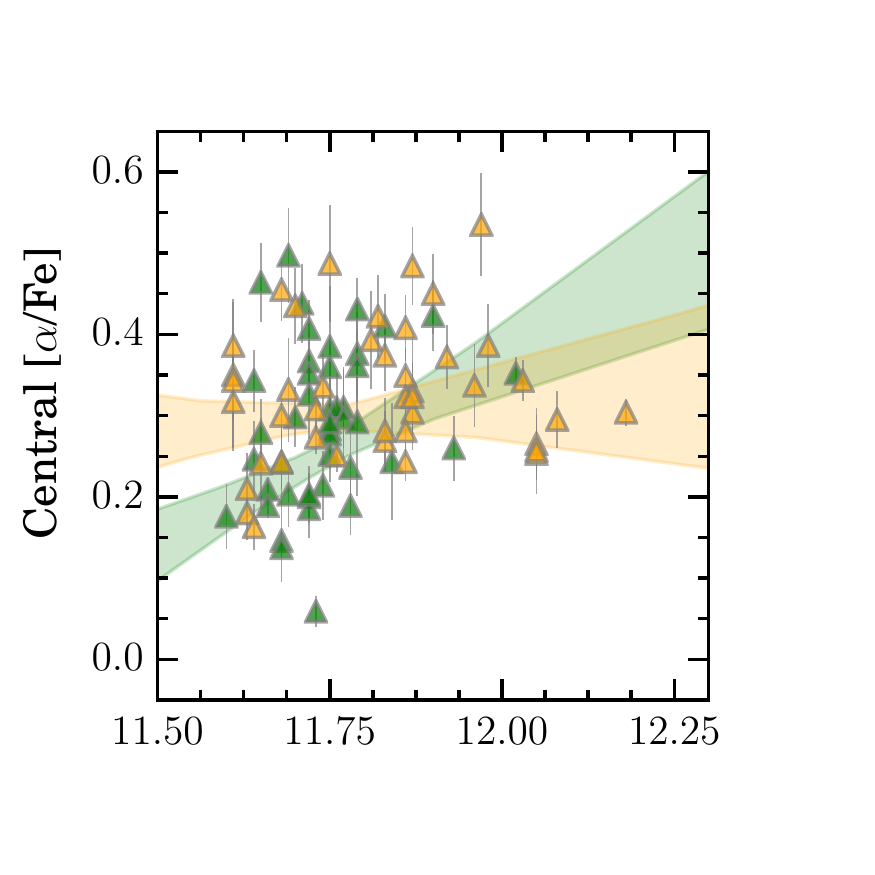}
}
\vskip -9mm
\vbox{
\includegraphics[width=0.8\columnwidth]{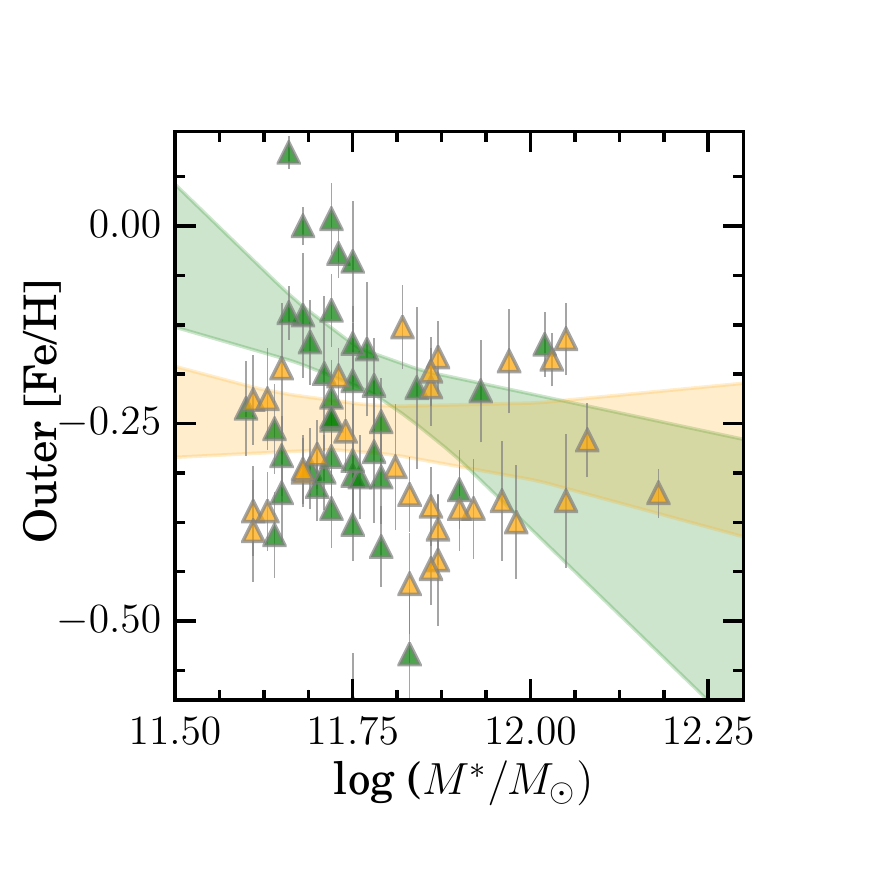}
\hskip -3mm
\includegraphics[width=0.8\columnwidth]{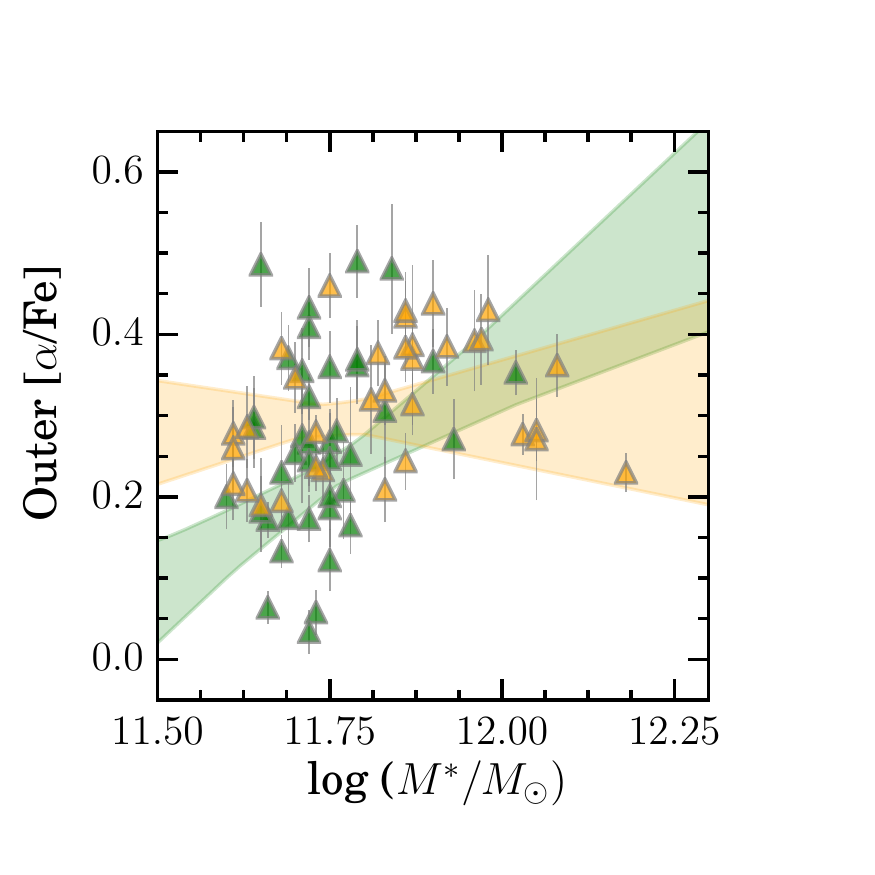}
}
\end{center}
\caption{We ask whether stellar populations know about large-scale 
environment $\delta_g$ at 
fixed \mstar. We fit a relation of the form 
$X = \alpha + \beta$~log($M^*/5.6 \times 10^{11}$~\msun), for $X=$\feh, \afe\ 
in two bins divided at log ($1+\delta_g$)$=0.6$, with low $\delta_g$ points (fits) 
in green and high $\delta_g$ points (fits) in yellow (see also Table 4). We see that the 
relations are significantly flatter in the high-density regions at fixed \mstar, mostly 
driven by galaxies at the lower-mass end.
}
\label{fig:afe-env}
\end{figure*}

\subsection{Correlations with Environment}
\label{sec:env}

We now look for trends between stellar populations and large-scale
($\delta_g$) and local ($\nu$) environmental measures (these are
reviewed in \S2.1). First we naively investigate correlations between
\feh\ and \afe\ (or gradients therein) with $\delta_g$ or $\nu$
without controlling for stellar mass. Table 3 contains all of the
correlations with $P<0.05$, again using the Spearman rank correlation
test.  No central stellar population parameters show any correlations
with environmental measures ($P>0.1$ in all cases), but we see
significant correlations between both environmental measures and
\feh\ and \afe\ at larger radius (Figure \ref{fig:env}). In general,
correlations with $\delta_g$ (the large-scale environment measure)
appear more significant than those with $\nu$, the local
indicator. The sense of the trend is that \feh\ is lower and \afe\ is
higher in overdense regions. This type of trend is as we would
expect if galaxies in the overdense regions formed earlier. We see a weak 
correlation between gradients in \afe\ and $\delta_g$, which we were not able 
to detect with our earlier stacking analysis in \citet{greeneetal2015}.

Given the strong covariance between stellar mass and environment
\citep[see review in][]{blantonmoustakas2009}, and the observed
correlation between \mstar\ and stellar population parameters (Figure
\ref{fig:afe-sig}), we must investigate whether the correlations
between stellar population parameters and environment are simply a
biproduct of a dominant correlation with
\mstar\ \citep[e.g.,][]{vealeetal2017a,vealeetal2017b}. We
  therefore fit a relation between \mstar\ and \feh\ or \afe, in two
  bins of $\delta_g$. We divide the sample at the median value of log
  ($1+\delta_g$)$=0.6$, so that there are even numbers of galaxies in
  each bin. The results are shown in Figure \ref{fig:afe-env} and the
  fits are presented in Table 4.

We find significant slope differences in the \mstar-\feh\ and
\mstar-\afe\ fits to the two density bins. Specifically, for both
\feh\ and \afe, the high-density systems show a much flatter relation
between \mstar\ and stellar population parameters than in the systems
in the lower-density environments. The slope differences are
significantly different in all cases, although they are larger in the
$0.9 R_e$ bin. As shown by the figure, at high mass (where we have
sparse data) the two density bins are convergent, while the different
slopes arise mainly because the lower-density points have higher
\feh\ and commensurately lower \afe\ than galaxies of similar mass in
higher-density environments. Similar trends have been reported by a
number of authors \citep[][and see \S
  \ref{sec:discussion}]{thomasetal2005,liuetal2016,guetal2018}. 
Specifically, the \atd\ stellar population data from
\citet{mcdermidetal2015} show a similar split in
\afe\ and \feh\ by local overdensity in their highest mass bin.

\begin{deluxetable}{cccll}
\tablenum{4}
\tablecolumns{6} 
\tabletypesize{\scriptsize}
\tablewidth{0pc}
\tablecaption{Mass-Environment Fits \label{tablecorr}}
\tablehead{ 
\colhead{Stellar Pop} & \colhead{Radius} & \colhead{Env.} & 
\colhead{$\alpha$} & \colhead{$\beta$} \\
\colhead{(1)} & \colhead{(2)} & \colhead{(3)} & 
\colhead{(4)} & \colhead{(5)}} 
\startdata
   \afe\ & $<2$kpc & all & $0.265 \pm 0.004$ & $0.23 \pm 0.03$ \\
   \afe\ & $<2$kpc & low & $0.255 \pm 0.005$ & $0.44 \pm 0.05$ \\
   \afe\ & $<2$kpc & high & $0.297 \pm 0.006$  & $0.08 \pm 0.03$  \\
   \afe\ &  $0.9R_e$ & all & $0.247 \pm 0.005$ & $0.26 \pm 0.03$ \\
   \afe\ &  $0.9R_e$ & low & $0.230 \pm 0.006$ & $0.53 \pm 0.07$ \\
   \afe\ &  $0.9R_e$ & high & $0.300 \pm 0.008$ & $0.05 \pm 0.04$ \\
\hline
   \feh\ & $<2$kpc & all & $-0.0618 \pm 0.005$ & $0.073 \pm 0.03$ \\
   \feh\ & $<2$kpc & low & $-0.0417 \pm 0.007$ & $-0.15 \pm 0.09$ \\
   \feh\ & $<2$kpc & high & $-0.111 \pm 0.008$  & $0.26 \pm 0.04$  \\
   \feh\ &  $0.9R_e$ & all & $-0.197 \pm 0.006$ & $-0.31 \pm 0.04$ \\
   \feh\ &  $0.9R_e$ & low & $-0.173 \pm 0.008$ & $-0.49 \pm 0.10$ \\
   \feh\ &  $0.9R_e$ & high & $-0.25 \pm 0.01$ & $-0.08 \pm 0.05$ \\
\enddata
\tablecomments{
We fit a relation of the form $X = \alpha + \beta$~log($M^*/5.6 \times 10^{11}$~\msun), 
for $X=$\feh, \afe. 
Col. (1): Stellar population parameter.
Col. (2): Radial extraction radius.
Col. (3): Environmental bin (all, low means log $(1+\delta_g) < 0.6$, high 
means log $(1+\delta_g) > 0.6$.
Col. (4): Zeropoint in the fit.
Col. (5): Slope of the fit.
Col. (6): Spearman probability.
}
\end{deluxetable}

We are looking at these trends binned in $\delta_g$, which looks at
overdensities on $\sim 5$~Mpc scales. Much of the literature has
looked at $\nu$, or local overdensity. Discussing the relative merits
of different environment indicators is beyond the scope of this work,
but we note here that $\nu$ in particular can be a complex
environmental indicator, since at low densities it is sensitive to
large-scale environment (the so-called ``two-halo'' term), while in
dense regions it probes small scales \citep[or the ``one-halo'' term;
  see ][]{wooetal2013}.  Accordingly, as shown by
\citet{vealeetal2017b}, $\nu$ and $\delta_g$ are well correlated at low
density (where $\nu$ probes $>5$ Mpc scales anyway) but diverge in
high density regions, where $\nu$ can grow much more rapidly. In our
case, given our limited numbers, the median $\delta_g$ value of
log$(1+\delta_g)\sim 1.5$ is close to the value where the two indicators
diverge, and thus we see very similar trends when we play the same
game with $\nu$.

\section{Discussion}
\label{sec:discussion}

\subsection{Ex-situ fractions, merger mass ratios, and expected stellar populations}

Motivated by observations of very compact sizes for quiescent galaxies
at high redshift \citep[e.g.,][]{vandokkumetal2008,vanderweletal2014},
theorists have examined the growth histories of massive galaxies and
determined that there is often a two-phase growth. An early and rapid
dissipational phase builds a compact central nugget, and is followed
by a build-up of the outer parts of galaxies via minor merging
\citep[e.g.,][]{oseretal2010,lackneretal2012,rodriguez-gomezetal2016,
  wellonsetal2016}. However, the relative roles of minor merging and
progenitor bias in the aggregate growth of the galaxy population
remains a matter of debate
\citep[e.g.,][]{newmanetal2012}. Observations of light profiles of
low-redshift early-type galaxies also provide some support for a
two-phase picture
\citep[e.g.,][]{huangetal2013,dsouzaetal2014,ohetal2017}, but adding
stellar population and kinematic information at large radius may add
additional insight.

A number of cosmological and cosmological zoom studies have looked at
how mass is accumulated in massive galaxies with time. They find that
the fraction of total stellar mass brought in via mergers rises
towards more massive galaxies
\citep[e.g.,][]{oseretal2010,lackneretal2012,hirschmannetal2015}.
Despite this overall trend, there is considerable scatter from system
to system because of different merger histories, as emphasized by
\citet{rodriguez-gomezetal2016}.

In terms of the radii that are most impacted by merging, the more
massive and more tightly bound satellites deposit their stars closer
to the center, while the lower-mass and less-bound systems deposit
mass at larger radius \citep[see
  also][]{boylankolchinma2007,amorisco2017}. Accordingly, major
mergers will tend to flatten the steep gradients that arise from pure
in-situ formation
\citep[e.g.,][]{white1980,kobayashi2004,hirschmannetal2015}. As the
dominant merger type becomes more minor, the accreted systems tend to
have their star formtion truncated early, when they fall into the
larger halo but before they are accreted into the primary
galaxy. Thus, minor mergers tend to create positive age gradients,
bring in $\alpha-$enhanced stars, and steepen metallicity gradients
\citep{hirschmannetal2015}.  This expectation is also consistent with
emerging observations of the stellar populations of massive galaxies
at $1<z<2$, which also show higher \afe\ and lower \feh\ in the
population that our MASSIVE galaxies may have later accreted
\citep[e.g.,][]{onoderaetal2015,lonoceetal2015,krieketal2016}.

In the context of these simulations, our goal is to ask whether
various structural measurements may link to the merger history, and
then whether the stellar populations show any correlations with
them. For instance, at the highest \mstar\ in our sample, we see that
the central \afe\ values match those in galaxies factors of $2-3$ less
massive. The most likely explanation is that major merging dominates
the mass growth at the centers of these galaxies. The trend between
\feh\ and gravitational potential also points to a significant contribution 
from in-situ stars $< R_e$, since those stars preserve a memory of the 
gravitational potential in which they formed.
Larger radius measurements are closer to the spatial regions
where more minor mergers would deposit their stars.

\subsection{Observable consequences of varying ex-situ fractions and links with 
stellar populations}

\begin{figure*}
\begin{center}
\vbox{
\includegraphics[width=0.75\columnwidth]{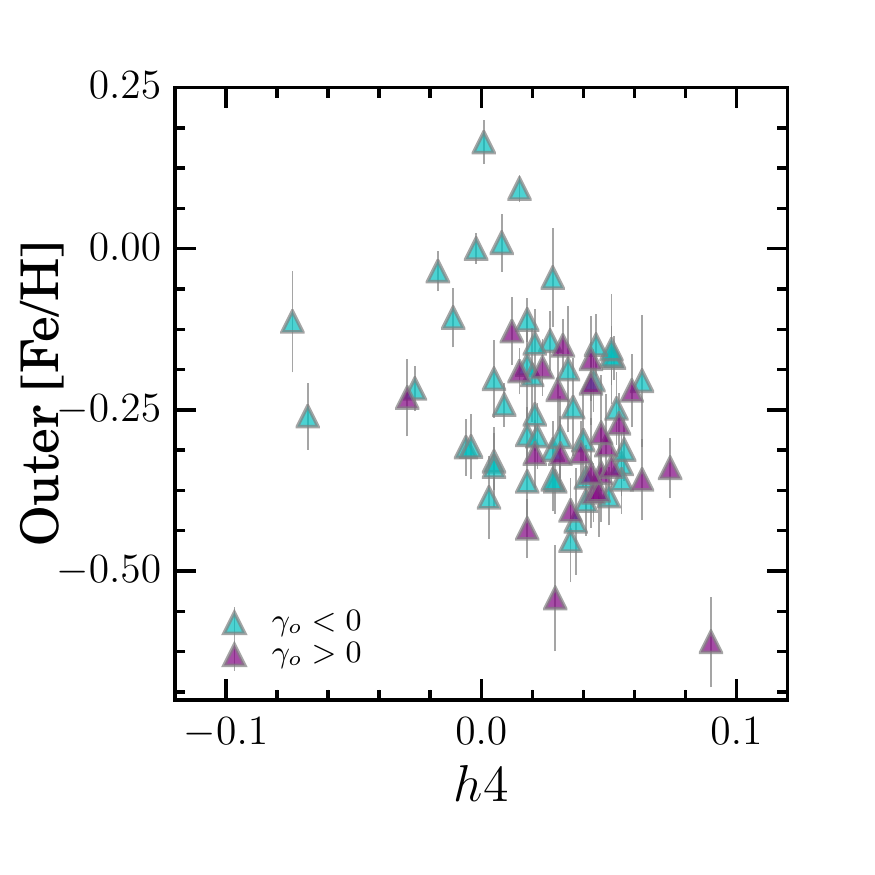}
\hskip -2mm
\includegraphics[width=0.75\columnwidth]{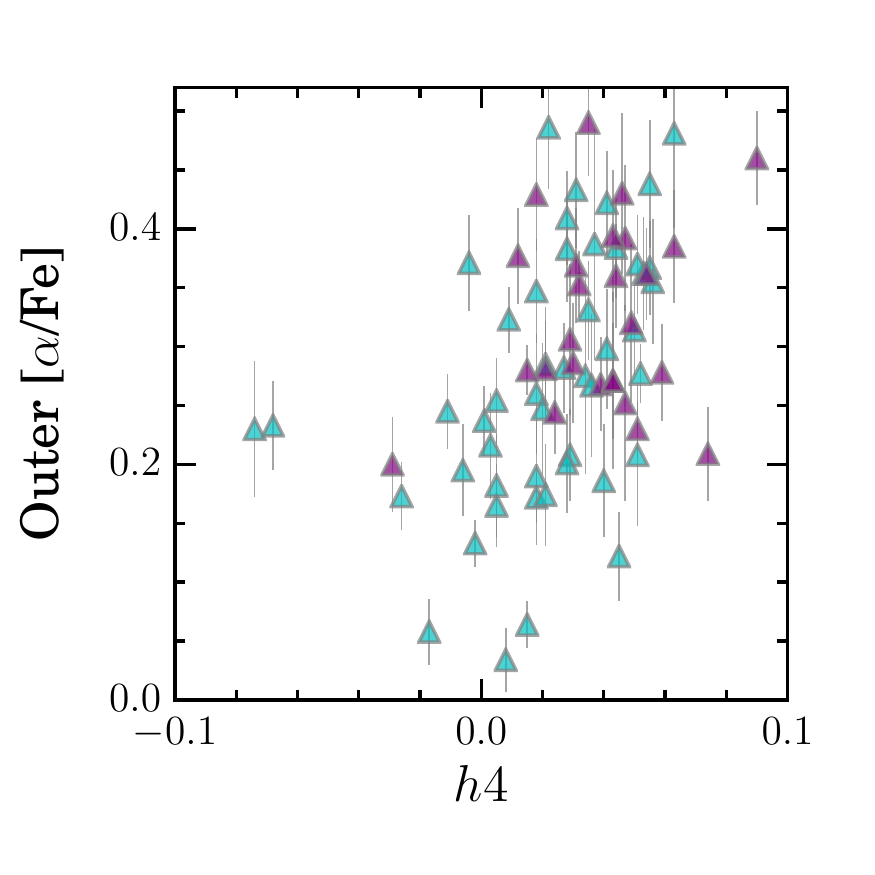}
}
\end{center}
\vskip -5mm
\caption{The relationship between $h4$ and \feh\ (left) and \afe\ (right) 
as measured at $0.9 R_e$, now split into galaxies with rising \sig\ (red 
circles) and falling \sig\ (blue triangles). 
All of the galaxies with rising \sig, and the galaxies with high 
$h4$ and falling \sig\ share similar stellar populations and carry the 
signature of increased accretion of satellites to grow their outer parts.
}
\label{fig:h4gamo}
\end{figure*}

We now consider how we might sort galaxies based on their accretion
histories. Stellar population gradients alone are insufficient, since
steep gradients can result from predominantly in-situ formation or a
large number of minor mergers \citep[e.g.,][]{kobayashi2004,hirschmannetal2015}.  We
have explored two structural characteristics here, the average $h4$
value \citep{thomasetal2007,wuetal2014,amorisco2017} and the outer surface brightness
slope $\Delta \Sigma_{23}$ \citep{pillepichetal2014,cooketal2016}. We
will discuss each of these in turn.

We see strong correlations between both \feh\ and \afe\ and $h4$. We
have argued that template mismatch is unlikely to drive this
correlation. We now ask whether the correlations are linked with the
accretion history, or instead some other property (e.g., the
gravitational potential) of the galaxy.  Simulations have shown that
radial anisotropy, and thus $h4$, can rise with merging
\citep{wuetal2014}. On the other hand, positive $h4$ can result from
gradients in circular velocity as well
\citep[e.g.,][]{gerhard1993,baesetal2005}.  In the MASSIVE sample,
\citet{vealeetal2017a,vealeetal2018} show that there are likely two
causes for positive $h4$, particularly in the outer parts of our
galaxies.  In two thirds of the galaxies, \sig\ falls outward. Those
galaxies with falling \sig\ and positive $h4$ likely have rising
radial anisotropy. While major mergers can produce radial orbits at
large radius \citep[e.g.,][]{rantalaetal2018}, N-body simulations
suggest that the stronger the radial anisotropy is, the more likely it
originates from minor mergers and/or accretion
\citep[][]{hilzetal2012}.  In the other third of galaxies (typically
the most massive), we see rising \sig\ profiles \citep[e.g.,][ and
  references therein]{dressler1979,loubseretal2008,vealeetal2018}. In
these galaxies, $h4$ will rise outward simply because of the gradient
in the potential.

We thus revisit the correlation between $h4$ and the outer stellar
populations, this time splitting the sample into those with rising and
falling \sig\ profiles (Fig.\ \ref{fig:h4gamo}). We find that the
galaxies with rising \sig\ profiles show uniformly high \afe\ and
relatively low \feh, while the galaxies with falling \sig\ profiles
span the full range of \afe\ and \feh, and these drive the correlation
with $h4$.  In these galaxies, we posit that the higher $h4$ comes
from radial anisotropy and that the outer stellar populations are
closely linked to the anisotropy of their orbits. The higher the
radial anisotropy, the more $\alpha$-enhanced (and thus older and
metal-poor) are the stellar populations. The range in radial anisotropy is linked 
to the merger history, perhaps pointing to more minor merging in galaxies 
with more radial anisotropy \citep[][]{hilzetal2012}.

Based on this behavior, we propose the following picture.  Galaxies
with falling \sig\ profiles display a sequence in ex-situ fraction,
whereby those with low $h4$ are also those with quiet accretion
histories, leading to solar-like abundances and abundance ratios in
their outskirts. Those with the highest $h4$ have high radial
anisotropy due to additional merging, which brings in older,
metal-poor and $\alpha-$enhanced stars.  Galaxies with rising
\sig\ profiles have high $h4$, but in these cases radial anisotropy
need not be invoked.  These also tend to be the galaxies in the most
massive halos \citep{vealeetal2018}, and we are likely seeing a
transition into an outer envelope or intra-cluster light component. 
The stellar populations that we measure in the outer parts of
these galaxies (namely very high \afe\ and low \feh) are consistent
with the stellar populations measured in the envelopes of brightest
cluster galaxies in other works
\citep[e.g.,][]{coccatoetal2011,edwardsetal2016}.  We will revisit
these galaxy outskirts from the perspective of environment in the next
section.

\citet{pillepichetal2014} suggest that gradients in surface brightness
profile (here we measure $\Delta \Sigma_{23}$) should be a robust way
to sort galaxies of similar mass by their ex-situ
fractions. \citet{cooketal2016} use the same Illustris simulations to
predict shallower \feh\ gradients in systems with the shallowest
surface brightness slopes.  We see a correlation between $\Delta
\Sigma_{23}$ and the \feh\ measured at $0.9 R_e$, but no correlation
with stellar population gradients. Thus, there is a hint that the gradient 
in surface brightness is linked with the stellar populations, and the correlation 
runs in a similar way as suggested by Cook et al.\ However, we have a few limitations 
currently to making a definitive comparison. Despite our efforts, we still cannot 
measure the stellar populations at matching radii to the surface brightness 
gradients. Furthermore, we would need a 
wider dynamic range in \mstar\ to determine whether 
$\Delta \Sigma_{23}$ correlates with stellar population properties.

\subsection{Environment and Galaxy Assembly}

We observe a hint of lower \feh\ and enhanced \afe\ at fixed mass in
more dense environments, both when environment is measured as local
overdensity ($\nu$) and when measured on Mpc scales ($\delta$). 
We can understand this result in the context of prior work focused both 
on the fossil record, looking at star formation histories of massive galaxies 
as a function of environment, and work that looks at the star formation 
rates in overdensities at different redshifts.

Previous work has seen clear evidence that star formation ended
earlier in denser environments
\citep[e.g.,][]{thomasetal2005,scottetal2017}, particularly at lower
\mstar\ where ``rejuvenated'' spheroids are only found in low-density
environments \citep[e.g.,][]{thomasetal2010,pasqualietal2010}. In
fact, \citet{liuetal2016} show that low-mass spheroids have a very
large range in \afe\ both in the densest and least dense environments. 
\citet{liuetal2016} also find that \afe\ in dwarfs correlates with 
the galaxy distance from M87, also consistent with our picture.
\citet{mcdermidetal2015} additionally show that the star-formation
histories of cluster galaxies are not just truncated earlier but
actually proceed more rapidly (e.g., the same stellar mass is built up
in a shorter burst) in denser environments, leading potentially to
lower \feh\ and higher \afe. Observations of star formation in dense
environments at $z>1.5$ also support the idea that cluster galaxies
experience more intense bursts of star formation \citep[e.g.,][]{elbazetal2007,
brodwinetal2013,tranetal2010,noirotetal2018}. 

Of particular interest to our study, 
\citet{guetal2018} find that all galaxies in the central parts 
of the Abell 3827 cluster show enhanced \afe\ for their mass. They argue that all 
galaxies in cluster centers have experienced rapid and early truncation of 
their star formation, and that this ``coordinated assembly'' of the satellite 
and central galaxy leads directly to the flat \afe\ gradients that they 
observe in the central galaxy. 

Gu et al.\ also suggest that the radial gradient in \afe\ should be
flatter in overdense environments, since all the galaxies available
for accreting will also contain elevated \afe\ for their mass. In
contrast, a relatively isolated MASSIVE galaxy might ingest a
satellite with ``normal'' \afe\ for its mass, leading to a declining
\afe\ gradient. Unlike in \citet{greeneetal2015} where we did not have the 
statistics, we do in fact see a weak trend between the
\afe\ gradients and the environment in just this sense.

\section{Summary}
\label{sec:conclusions}

We look at the stellar populations throughout 90 MASSIVE galaxies,
with a focus on the outer parts of the galaxies. In moving beyond measurements 
of stellar populations weighted towards the galaxy centers, we are able 
to uncover new trends between stellar population parameters and the 
kinematics of these massive galaxies, providing new insight into their 
assembly history.

In galaxy centers, we see that the \afe\ saturates at stellar masses
$\sim 10^{11.8}$~\msun, suggesting that major mergers play a large
role in building up the centers of the most massive galaxies.  At and beyond $\sim
R_e$, \sig\ and \afe\ are tightly correlated. In contrast, the
metallicity \feh\ is correlated with the gravitational potential and
the surface mass density, pointing to a significant in-situ component
to the stars at these radii. 

We investigate two structural measurements that may correlate with the
ex-situ (or accreted) fraction.  We find a correlation between the
outer surface brightness slope $\Delta \Sigma_{23}$ and \feh\ beyond
$\sim R_e$. Our finding provides some support for the Illustris
results from \citet{pillepichetal2014} and \citet{cooketal2016} that
the outer surface brightness slope is a proxy for accreted or ex-situ
fraction. To truly compare with simulations, a wider dynamic range in
stellar mass is needed.

We also find a strong correlation between \afe\ and $h4$, and we even
see a correlation between \feh\ and $h4$ at $\sim R_e$. Galaxies with
the most positive $h4$ have super-solar \afe$\approx 0.4-0.6$ dex
and low \feh$\approx -0.5$ dex.  Mergers can increase the radial
anisotropy and thus boost $h4$, or $h4$ can rise due to a gradient in
the potential (e.g., the transition from a galaxy halo to a cluster
halo). Even removing the galaxies with rising \sig\ profiles, we still
find a strong correlation with stellar populations, pointing to a
rising accreted population in galaxies with higher radial anisotropy.

The galaxies with rising \sig\ profiles at large radius also tend to be those 
in the densest environments \citep{vealeetal2018}, and 
we find evidence that the \afe\ in the outer parts correlates with 
environment at fixed mass. Lower-density environments have lower \afe,
while higher-density environments have very super-solar \afe$\sim
0.4-0.6$. If star formation proceeds earlier and more rapidly in denser 
environments, then we would see older, metal-poorer, and more $\alpha-$enhanced 
stars in the outskirts of galaxies in richer environments.

We are interested in probing the assembly history of massive galaxies
by looking at their stellar populations at large radius. Although our
survey was designed to reach large radii for very massive galaxies,
and does as well as any existing data set, there is some real chance
that our observations still do not reach out to large enough radius to
truly constrain the accreted stars. Current imaging surveys are now able to 
reach out to $\sim 100$ kpc for individual galaxies with photometry 
\citep[e.g.,][]{huangetal2018a}, but we must wait for next-generation 
telescopes to have the collecting area to measure detailed stellar population 
properties out to comparable radii. 



\section*{Acknowledgements}

We thank the referee for a timely, supportive, but thoughtful report that improved 
this manuscript. The MASSIVE survey is supported in part by NSF AST-1411945, AST-1411642, AST-1815417, AST-1817100,  HST GO-14210, GO-15265 and AR-14573.  We are grateful for useful discussions with M.\ Kriek.



\begin{thebibliography}{}
\expandafter\ifx\csname natexlab\endcsname\relax\def\natexlab#1{#1}\fi
\providecommand{\url}[1]{\href{#1}{#1}}
\providecommand{\dodoi}[1]{doi:~\href{http://doi.org/#1}{\nolinkurl{#1}}}
\providecommand{\doeprint}[1]{\href{http://ascl.net/#1}{\nolinkurl{http://ascl.net/#1}}}
\providecommand{\doarXiv}[1]{\href{https://arxiv.org/abs/#1}{\nolinkurl{https://arxiv.org/abs/#1}}}

\bibitem[{{Adams} {et~al.}(2011){Adams}, {Blanc}, {Hill}, {Gebhardt}, {Drory},
  {Hao}, {Bender}, {Byun}, {Ciardullo}, {Cornell}, {Finkelstein}, {Fry},
  {Gawiser}, {Gronwall}, {Hopp}, {Jeong}, {Kelz}, {Kelzenberg}, {Komatsu},
  {MacQueen}, {Murphy}, {Odoms}, {Roth}, {Schneider}, {Tufts}, \&
  {Wilkinson}}]{adamsetal2011}
{Adams}, J.~J., {Blanc}, G.~A., {Hill}, G.~J., {et~al.} 2011, \apjs, 192, 5,
  \dodoi{10.1088/0067-0049/192/1/5}

\bibitem[{{Amorisco}(2017)}]{amorisco2017}
{Amorisco}, N.~C. 2017, \mnras, 464, 2882, \dodoi{10.1093/mnras/stw2229}

\bibitem[{{Annibali} {et~al.}(2007){Annibali}, {Bressan}, {Rampazzo},
  {Zeilinger}, \& {Danese}}]{annibalietal2007}
{Annibali}, F., {Bressan}, A., {Rampazzo}, R., {Zeilinger}, W.~W., \& {Danese},
  L. 2007, \aap, 463, 455, \dodoi{10.1051/0004-6361:20054726}

\bibitem[{{Baes} {et~al.}(2005){Baes}, {Dejonghe}, \& {Buyle}}]{baesetal2005}
{Baes}, M., {Dejonghe}, H., \& {Buyle}, P. 2005, \aap, 432, 411,
  \dodoi{10.1051/0004-6361:20041907}

\bibitem[{{Barone} {et~al.}(2018){Barone}, {D'Eugenio}, {Colless},
  {Scott}, {van de Sande}, {Bland-Hawthorn}, {Brough}, {Bryant}, {Cortese},
  {Croom}, {Foster}, {Goodwin}, {Konstantopoulos}, {Lawrence}, {Lorente},
  {Medling}, {Owers}, \& {Richards}}]{baroneetal2018}
{Barone}, T.~M., {D'Eugenio}, F., {Colless}, M., {et~al.} 2018, \apj,
  856, 64, \dodoi{10.3847/1538-4357/aaaf6e}

\bibitem[{{Barro} {et~al.}(2013){Barro}, {Faber}, {P{\'e}rez-Gonz{\'a}lez},
  {Koo}, {Williams}, {Kocevski}, {Trump}, {Mozena}, {McGrath}, {van der Wel},
  {Wuyts}, {Bell}, {Croton}, {Ceverino}, {Dekel}, {Ashby}, {Cheung},
  {Ferguson}, {Fontana}, {Fang}, {Giavalisco}, {Grogin}, {Guo}, {Hathi},
  {Hopkins}, {Huang}, {Koekemoer}, {Kartaltepe}, {Lee}, {Newman}, {Porter},
  {Primack}, {Ryan}, {Rosario}, {Somerville}, {Salvato}, \&
  {Hsu}}]{barroetal2013}
{Barro}, G., {Faber}, S.~M., {P{\'e}rez-Gonz{\'a}lez}, P.~G., {et~al.} 2013,
  \apj, 765, 104, \dodoi{10.1088/0004-637X/765/2/104}

\bibitem[{{Bezanson} {et~al.}(2009){Bezanson}, {van Dokkum}, {Tal},
  {Marchesini}, {Kriek}, {Franx}, \& {Coppi}}]{bezansonetal2009}
{Bezanson}, R., {van Dokkum}, P.~G., {Tal}, T., {et~al.} 2009, \apj, 697, 1290,
  \dodoi{10.1088/0004-637X/697/2/1290}

\bibitem[{{Blakeslee}(2013)}]{blakeslee2013}
{Blakeslee}, J.~P. 2013, in IAU Symposium, Vol. 289, Advancing the Physics of
  Cosmic Distances, ed. R.~{de Grijs}, 304--311

\bibitem[{{Blakeslee} {et~al.}(2009){Blakeslee}, {Jord{\'a}n}, {Mei},
  {C{\^o}t{\'e}}, {Ferrarese}, {Infante}, {Peng}, {Tonry}, \&
  {West}}]{blakesleeetal2009}
{Blakeslee}, J.~P., {Jord{\'a}n}, A., {Mei}, S., {et~al.} 2009, \apj, 694, 556,
  \dodoi{10.1088/0004-637X/694/1/556}

\bibitem[{{Blakeslee} {et~al.}(2010){Blakeslee}, {Cantiello}, {Mei},
  {C{\^o}t{\'e}}, {Barber DeGraaff}, {Ferrarese}, {Jord{\'a}n}, {Peng},
  {Tonry}, \& {Worthey}}]{blakesleeetal2010}
{Blakeslee}, J.~P., {Cantiello}, M., {Mei}, S., {et~al.} 2010, \apj, 724, 657,
  \dodoi{10.1088/0004-637X/724/1/657}

\bibitem[{{Blanton} \& {Moustakas}(2009)}]{blantonmoustakas2009}
{Blanton}, M.~R., \& {Moustakas}, J. 2009, \araa, 47, 159,
  \dodoi{10.1146/annurev-astro-082708-101734}

\bibitem[{{Boardman} {et~al.}(2017){Boardman}, {Weijmans}, {van den Bosch},
  {Kuntschner}, {Emsellem}, {Cappellari}, {de Zeeuw}, {Falc{\'o}n-Barroso},
  {Krajnovi{\'c}}, {McDermid}, {Naab}, {van de Ven}, \&
  {Yildirim}}]{boardmanetal2017}
{Boardman}, N.~F., {Weijmans}, A.-M., {van den Bosch}, R., {et~al.} 2017,
  \mnras, 471, 4005, \dodoi{10.1093/mnras/stx1835}

\bibitem[{{Boylan-Kolchin} \& {Ma}(2007)}]{boylankolchinma2007}
{Boylan-Kolchin}, M., \& {Ma}, C.-P. 2007, \mnras, 374, 1227,
  \dodoi{10.1111/j.1365-2966.2006.11276.x}

\bibitem[{{Brodwin} {et~al.}(2013){Brodwin}, {Stanford}, {Gonzalez}, {Zeimann},
  {Snyder}, {Mancone}, {Pope}, {Eisenhardt}, {Stern}, {Alberts}, {Ashby},
  {Brown}, {Chary}, {Dey}, {Galametz}, {Gettings}, {Jannuzi}, {Miller},
  {Moustakas}, \& {Moustakas}}]{brodwinetal2013}
{Brodwin}, M., {Stanford}, S.~A., {Gonzalez}, A.~H., {et~al.} 2013, \apj, 779,
  138, \dodoi{10.1088/0004-637X/779/2/138}

\bibitem[{{Burstein}(1985)}]{burstein1985}
{Burstein}, D. 1985, \pasp, 97, 89, \dodoi{10.1086/131502}

\bibitem[{{Cappellari}(2013)}]{cappellari2013}
{Cappellari}, M. 2013, \apjl, 778, L2, \dodoi{10.1088/2041-8205/778/1/L2}

\bibitem[{{Cappellari} \& {Emsellem} (2004)}]{cappellariemsellem2004}
Cappellari, M., \& Emsellem, E.\ 2004, \pasp, 116, 138, \dodoi{10.1086/381875}

\bibitem[{{Carrick} {et~al.}(2015){Carrick}, {Turnbull}, {Lavaux}, \&
  {Hudson}}]{carricketal2015}
{Carrick}, J., {Turnbull}, S.~J., {Lavaux}, G., \& {Hudson}, M.~J. 2015,
  \mnras, 450, 317, \dodoi{10.1093/mnras/stv547}

\bibitem[{{Coccato} {et~al.}(2011){Coccato}, {Gerhard}, {Arnaboldi}, \&
  {Ventimiglia}}]{coccatoetal2011}
{Coccato}, L., {Gerhard}, O., {Arnaboldi}, M., \& {Ventimiglia}, G. 2011, \aap,
  533, A138, \dodoi{10.1051/0004-6361/201117546}

\bibitem[{{Conroy} {et~al.}(2014){Conroy}, {Graves}, \& {van
  Dokkum}}]{conroyetal2014}
{Conroy}, C., {Graves}, G.~J., \& {van Dokkum}, P.~G. 2014, \apj, 780, 33,
  \dodoi{10.1088/0004-637X/780/1/33}

\bibitem[{{Cook} {et~al.}(2016){Cook}, {Conroy}, {Pillepich},
  {Rodriguez-Gomez}, \& {Hernquist}}]{cooketal2016}
{Cook}, B.~A., {Conroy}, C., {Pillepich}, A., {Rodriguez-Gomez}, V., \&
  {Hernquist}, L. 2016, \apj, 833, 158, \dodoi{10.3847/1538-4357/833/2/158}

\bibitem[{{Crook} {et~al.}(2007){Crook}, {Huchra}, {Martimbeau}, {Masters},
  {Jarrett}, \& {Macri}}]{crooketal2007}
{Crook}, A.~C., {Huchra}, J.~P., {Martimbeau}, N., {et~al.} 2007, \apj, 655,
  790, \dodoi{10.1086/510201}

\bibitem[{{Dressler}(1979)}]{dressler1979}
{Dressler}, A. 1979, \apj, 231, 659, \dodoi{10.1086/157229}

\bibitem[{{D'Souza} {et~al.}(2014){D'Souza}, {Kauffman}, {Wang}, \&
  {Vegetti}}]{dsouzaetal2014}
{D'Souza}, R., {Kauffman}, G., {Wang}, J., \& {Vegetti}, S. 2014, \mnras, 443,
  1433, \dodoi{10.1093/mnras/stu1194}

\bibitem[{{Edwards} {et~al.}(2016){Edwards}, {Alpert}, {Trierweiler},
  {Abraham}, \& {Beizer}}]{edwardsetal2016}
{Edwards}, L.~O.~V., {Alpert}, H.~S., {Trierweiler}, I.~L., {Abraham}, T., \&
  {Beizer}, V.~G. 2016, \mnras, 461, 230, \dodoi{10.1093/mnras/stw1314}

\bibitem[{{Elbaz} {et~al.}(2007){Elbaz}, {Daddi}, {Le Borgne}, {Dickinson},
  {Alexander}, {Chary}, {Starck}, {Brandt}, {Kitzbichler}, {MacDonald},
  {Nonino}, {Popesso}, {Stern}, \& {Vanzella}}]{elbazetal2007}
{Elbaz}, D., {Daddi}, E., {Le Borgne}, D., {et~al.} 2007, \aap, 468, 33,
  \dodoi{10.1051/0004-6361:20077525}

\bibitem[{{Emsellem} {et~al.}(2007)}]{emsellemetal2007}
{Emsellem}, E., {et~al.} 2007, \mnras, 379, 401,
  \dodoi{10.1111/j.1365-2966.2007.11752.x}

\bibitem[{{Ene} {et~al.}(2018){Ene}, {Ma}, {Veale}, {Greene}, {Thomas},
  {Blakeslee}, {Foster}, {Walsh}, {Ito}, \& {Goulding}}]{eneetal2018}
{Ene}, I., {Ma}, C.-P., {Veale}, M., {et~al.} 2018, \mnras, 479, 2810,
  \dodoi{10.1093/mnras/sty1649}

\bibitem[{{Faber} {et~al.}(1985){Faber}, {Friel}, {Burstein}, \&
  {Gaskell}}]{faberetal1985}
{Faber}, S.~M., {Friel}, E.~D., {Burstein}, D., \& {Gaskell}, C.~M. 1985,
  \apjs, 57, 711, \dodoi{10.1086/191024}

\bibitem[{{Gerhard}(1993)}]{gerhard1993}
{Gerhard}, O.~E. 1993, \mnras, 265, 213, \dodoi{10.1093/mnras/265.1.213}

\bibitem[{{Goddard} {et~al.}(2017){Goddard}, {Thomas}, {Maraston}, {Westfall},
  {Etherington}, {Riffel}, {Mallmann}, {Zheng}, {Argudo-Fern{\'a}ndez}, {Lian},
  {Bershady}, {Bundy}, {Drory}, {Law}, {Yan}, {Wake}, {Weijmans}, {Bizyaev},
  {Brownstein}, {Lane}, {Maiolino}, {Masters}, {Merrifield}, {Nitschelm},
  {Pan}, {Roman-Lopes}, {Storchi-Bergmann}, \& {Schneider}}]{goddardetal2017}
{Goddard}, D., {Thomas}, D., {Maraston}, C., {et~al.} 2017, \mnras, 466, 4731,
  \dodoi{10.1093/mnras/stw3371}

\bibitem[{{Goullaud} {et~al.}(2018){Goullaud}, {Jensen}, {Blakeslee}, {Ma},
  {Greene}, \& {Thomas}}]{goullaudetal2018}
{Goullaud}, C.~F., {Jensen}, J.~B., {Blakeslee}, J.~P., {et~al.} 2018, \apj,
  856, 11, \dodoi{10.3847/1538-4357/aab1f3}

\bibitem[{{Graves} {et~al.}(2009){Graves}, {Faber}, \&
  {Schiavon}}]{gravesetal2009}
{Graves}, G.~J., {Faber}, S.~M., \& {Schiavon}, R.~P. 2009, \apj, 698, 1590,
  \dodoi{10.1088/0004-637X/698/2/1590}

\bibitem[{{Graves} \& {Schiavon}(2008)}]{gravesschiavon2008}
{Graves}, G.~J., \& {Schiavon}, R.~P. 2008, \apjs, 177, 446,
  \dodoi{10.1086/588097}

\bibitem[{{Greene} {et~al.}(2015){Greene}, {Janish}, {Ma}, {McConnell},
  {Blakeslee}, {Thomas}, \& {Murphy}}]{greeneetal2015}
{Greene}, J.~E., {Janish}, R., {Ma}, C.-P., {et~al.} 2015, \apj, 807, 11,
  \dodoi{10.1088/0004-637X/807/1/11}

\bibitem[{{Greene} {et~al.}(2012){Greene}, {Murphy}, {Comerford}, {Gebhardt},
  \& {Adams}}]{greeneetal2012}
{Greene}, J.~E., {Murphy}, J.~D., {Comerford}, J.~M., {Gebhardt}, K., \&
  {Adams}, J.~J. 2012, \apj, 750, 32, \dodoi{10.1088/0004-637X/750/1/32}

\bibitem[{{Greene} {et~al.}(2013){Greene}, {Murphy}, {Graves}, {Gunn},
  {Raskutti}, {Comerford}, \& {Gebhardt}}]{greeneetal2013}
{Greene}, J.~E., {Murphy}, J.~D., {Graves}, G.~J., {et~al.} 2013, \apj, 776,
  64, \dodoi{10.1088/0004-637X/776/2/64}

\bibitem[{{Gu} {et~al.}(2018){Gu}, {Conroy}, \& {Brammer}}]{guetal2018}
{Gu}, M., {Conroy}, C., \& {Brammer}, G. 2018, ArXiv e-prints.
\newblock \doarXiv{1805.04520}

\bibitem[{{Hill} {et~al.}(2008)}]{hilletal2008}
{Hill}, G.~J., {et~al.} 2008, in Society of Photo-Optical Instrumentation
  Engineers (SPIE) Conference Series, Vol. 7014, Society of Photo-Optical
  Instrumentation Engineers (SPIE) Conference Series

\bibitem[{{Hilz} {et~al.}(2012){Hilz}, {Naab}, {Ostriker}, {Thomas}, {Burkert},
  \& {Jesseit}}]{hilzetal2012}
{Hilz}, M., {Naab}, T., {Ostriker}, J.~P., {et~al.} 2012, \mnras, 425, 3119,
  \dodoi{10.1111/j.1365-2966.2012.21541.x}

\bibitem[{{Hirschmann} {et~al.}(2015){Hirschmann}, {Naab}, {Ostriker},
  {Forbes}, {Duc}, {Dav{\'e}}, {Oser}, \& {Karabal}}]{hirschmannetal2015}
{Hirschmann}, M., {Naab}, T., {Ostriker}, J.~P., {et~al.} 2015, \mnras, 449,
  528, \dodoi{10.1093/mnras/stv274}

\bibitem[{{Huang} {et~al.}(2013){Huang}, {Ho}, {Peng}, {Li}, \&
  {Barth}}]{huangetal2013}
{Huang}, S., {Ho}, L.~C., {Peng}, C.~Y., {Li}, Z.-Y., \& {Barth}, A.~J. 2013,
  \apj, 766, 47, \dodoi{10.1088/0004-637X/766/1/47}

\bibitem[{{Huang} {et~al.}(2018{\natexlab{a}}){Huang}, {Leauthaud}, {Greene},
  {Bundy}, {Lin}, {Tanaka}, {Miyazaki}, \& {Komiyama}}]{huangetal2018b}
{Huang}, S., {Leauthaud}, A., {Greene}, J.~E., {et~al.} 2018{\natexlab{a}},
  \mnras, 475, 3348, \dodoi{10.1093/mnras/stx3200}

\bibitem[{{Huang} {et~al.}(2018{\natexlab{b}}){Huang}, {Leauthaud}, {Greene},
  {Bundy}, {Lin}, {Tanaka}, {Mandelbaum}, {Miyazaki}, \&
  {Komiyama}}]{huangetal2018a}
{Huang}, S., {Leauthaud}, A., {Greene}, J., {et~al.} 2018{\natexlab{b}}, ArXiv
  e-prints.
\newblock \doarXiv{1803.02824}

\bibitem[{{Jarrett} {et~al.}(2003){Jarrett}, {Chester}, {Cutri}, {Schneider},
  \& {Huchra}}]{jarrettetal2003}
{Jarrett}, T.~H., {Chester}, T., {Cutri}, R., {Schneider}, S.~E., \& {Huchra},
  J.~P. 2003, \aj, 125, 525, \dodoi{10.1086/345794}

\bibitem[{{Jimmy} {et~al.}(2013){Jimmy}, {Tran}, {Brough}, {Gebhardt}, {von der
  Linden}, {Couch}, \& {Sharp}}]{jimmyetal2013}
{Jimmy}, {Tran}, K.-V., {Brough}, S., {et~al.} 2013, \apj, 778, 171,
  \dodoi{10.1088/0004-637X/778/2/171}

\bibitem[{{Johansson} {et~al.}(2012){Johansson}, {Thomas}, \&
  {Maraston}}]{johanssonetal2012}
{Johansson}, J., {Thomas}, D., \& {Maraston}, C. 2012, \mnras, 421, 1908,
  \dodoi{10.1111/j.1365-2966.2011.20316.x}

\bibitem[{{Kobayashi}(2004)}]{kobayashi2004}
{Kobayashi}, C. 2004, \mnras, 347, 740,
  \dodoi{10.1111/j.1365-2966.2004.07258.x}

\bibitem[{{Kormendy} {et~al.}(2009){Kormendy}, {Fisher}, {Cornell}, \&
  {Bender}}]{kormendyetal2009}
{Kormendy}, J., {Fisher}, D.~B., {Cornell}, M.~E., \& {Bender}, R. 2009, \apjs,
  182, 216, \dodoi{10.1088/0067-0049/182/1/216}

\bibitem[{{Korn} {et~al.}(2005){Korn}, {Maraston}, \& {Thomas}}]{kornetal2005}
{Korn}, A.~J., {Maraston}, C., \& {Thomas}, D. 2005, \aap, 438, 685,
  \dodoi{10.1051/0004-6361:20042126}

\bibitem[{{Kriek} {et~al.}(2016){Kriek}, {Conroy}, {van Dokkum}, {Shapley},
  {Choi}, {Reddy}, {Siana}, {van de Voort}, {Coil}, \&
  {Mobasher}}]{krieketal2016}
{Kriek}, M., {Conroy}, C., {van Dokkum}, P.~G., {et~al.} 2016, \nat, 540, 248,
  \dodoi{10.1038/nature20570}

\bibitem[{{Lackner} {et~al.}(2012){Lackner}, {Cen}, {Ostriker}, \&
  {Joung}}]{lackneretal2012}
{Lackner}, C.~N., {Cen}, R., {Ostriker}, J.~P., \& {Joung}, M.~R. 2012, \mnras,
  425, 641, \dodoi{10.1111/j.1365-2966.2012.21525.x}

\bibitem[{{Lavaux} \& {Hudson}(2011)}]{lavauxhudson2011}
{Lavaux}, G., \& {Hudson}, M.~J. 2011, \mnras, 416, 2840,
  \dodoi{10.1111/j.1365-2966.2011.19233.x}

\bibitem[{{Liu} {et~al.}(2016){Liu}, {Peng}, {Blakeslee}, {C{\^o}t{\'e}},
  {Ferrarese}, {Jord{\'a}n}, {Puzia}, {Toloba}, \& {Zhang}}]{liuetal2016}
{Liu}, Y., {Peng}, E.~W., {Blakeslee}, J., {et~al.} 2016, \apj, 818, 179,
  \dodoi{10.3847/0004-637X/818/2/179}

\bibitem[{{Lonoce} {et~al.}(2015){Lonoce}, {Longhetti}, {Maraston}, {Thomas},
  {Mancini}, {Cimatti}, {Ciocca}, {Citro}, {Daddi}, {di Serego Alighieri},
  {Gargiulo}, {Maiolino}, {Mannucci}, {Moresco}, {Pozzetti}, {Quai}, \&
  {Saracco}}]{lonoceetal2015}
{Lonoce}, I., {Longhetti}, M., {Maraston}, C., {et~al.} 2015, \mnras, 454,
  3912, \dodoi{10.1093/mnras/stv2150}

\bibitem[{{Loubser} {et~al.}(2008){Loubser}, {Sansom},
  {S{\'a}nchez-Bl{\'a}zquez}, {Soechting}, \& {Bromage}}]{loubseretal2008}
{Loubser}, S.~I., {Sansom}, A.~E., {S{\'a}nchez-Bl{\'a}zquez}, P., {Soechting},
  I.~K., \& {Bromage}, G.~E. 2008, \mnras, 391, 1009,
  \dodoi{10.1111/j.1365-2966.2008.13813.x}

\bibitem[{{Ma} {et~al.}(2014){Ma}, {Greene}, {McConnell}, {Janish},
  {Blakeslee}, {Thomas}, \& {Murphy}}]{maetal2014}
{Ma}, C.-P., {Greene}, J.~E., {McConnell}, N., {et~al.} 2014, \apj, 795, 158,
  \dodoi{10.1088/0004-637X/795/2/158}

\bibitem[{{McDermid} {et~al.}(2015){McDermid}, {Alatalo}, {Blitz}, {Bournaud},
  {Bureau}, {Cappellari}, {Crocker}, {Davies}, {Davis}, {de Zeeuw}, {Duc},
  {Emsellem}, {Khochfar}, {Krajnovic}, {Kuntschner}, {Morganti}, {Naab},
  {Oosterloo}, {Sarzi}, {Scott}, {Serra}, {Weijmans}, \&
  {Young}}]{mcdermidetal2015}
{McDermid}, R.~M., {Alatalo}, K., {Blitz}, L., {et~al.} 2015, \mnras, accepted
  (arXiv:1501.03723).
\newblock \doarXiv{1501.03723}

\bibitem[{{Mehlert} {et~al.}(2003){Mehlert}, {Thomas}, {Saglia}, {Bender}, \&
  {Wegner}}]{mehlertetal2003}
{Mehlert}, D., {Thomas}, D., {Saglia}, R.~P., {Bender}, R., \& {Wegner}, G.
  2003, \aap, 407, 423, \dodoi{10.1051/0004-6361:20030886}

\bibitem[{{Mould} {et~al.}(2000){Mould}, {Huchra}, {Freedman}, {Kennicutt},
  {Ferrarese}, {Ford}, {Gibson}, {Graham}, {Hughes}, {Illingworth}, {Kelson},
  {Macri}, {Madore}, {Sakai}, {Sebo}, {Silbermann}, \&
  {Stetson}}]{mouldetal2000}
{Mould}, J.~R., {Huchra}, J.~P., {Freedman}, W.~L., {et~al.} 2000, \apj, 529,
  786, \dodoi{10.1086/308304}

\bibitem[{{Murphy} {et~al.}(2011){Murphy}, {Gebhardt}, \&
  {Adams}}]{murphyetal2011}
{Murphy}, J.~D., {Gebhardt}, K., \& {Adams}, J.~J. 2011, \apj, 729, 129,
  \dodoi{10.1088/0004-637X/729/2/129}

\bibitem[{{Naab} {et~al.}(2009){Naab}, {Johansson}, \&
  {Ostriker}}]{naabetal2009}
{Naab}, T., {Johansson}, P.~H., \& {Ostriker}, J.~P. 2009, \apjl, 699, L178,
  \dodoi{10.1088/0004-637X/699/2/L178}

\bibitem[{{Newman} {et~al.}(2012){Newman}, {Ellis}, {Bundy}, \&
  {Treu}}]{newmanetal2012}
{Newman}, A.~B., {Ellis}, R.~S., {Bundy}, K., \& {Treu}, T. 2012, \apj, 746,
  162, \dodoi{10.1088/0004-637X/746/2/162}

\bibitem[{{Noirot} {et~al.}(2018){Noirot}, {Stern}, {Mei}, {Wylezalek},
  {Cooke}, {De Breuck}, {Galametz}, {Hatch}, {Vernet}, {Brodwin}, {Eisenhardt},
  {Gonzalez}, {Jarvis}, {Rettura}, {Seymour}, \& {Stanford}}]{noirotetal2018}
{Noirot}, G., {Stern}, D., {Mei}, S., {et~al.} 2018, \apj, 859, 38,
  \dodoi{10.3847/1538-4357/aabadb}

\bibitem[{{Oh} {et~al.}(2017){Oh}, {Greene}, \& {Lackner}}]{ohetal2017}
{Oh}, S., {Greene}, J.~E., \& {Lackner}, C.~N. 2017, \apj, 836, 115,
  \dodoi{10.3847/1538-4357/836/1/115}

\bibitem[{{Onodera} {et~al.}(2015){Onodera}, {Carollo}, {Renzini},
  {Cappellari}, {Mancini}, {Arimoto}, {Daddi}, {Gobat}, {Strazzullo},
  {Tacchella}, \& {Yamada}}]{onoderaetal2015}
{Onodera}, M., {Carollo}, C.~M., {Renzini}, A., {et~al.} 2015, \apj, 808, 161,
  \dodoi{10.1088/0004-637X/808/2/161}

\bibitem[{{Oser} {et~al.}(2010){Oser}, {Ostriker}, {Naab}, {Johansson}, \&
  {Burkert}}]{oseretal2010}
{Oser}, L., {Ostriker}, J.~P., {Naab}, T., {Johansson}, P.~H., \& {Burkert}, A.
  2010, \apj, 725, 2312, \dodoi{10.1088/0004-637X/725/2/2312}

\bibitem[{{Pasquali} {et~al.}(2010){Pasquali}, {Gallazzi}, {Fontanot}, {van den
  Bosch}, {De Lucia}, {Mo}, \& {Yang}}]{pasqualietal2010}
{Pasquali}, A., {Gallazzi}, A., {Fontanot}, F., {et~al.} 2010, \mnras, 407,
  937, \dodoi{10.1111/j.1365-2966.2010.17074.x}

\bibitem[{{Pillepich} {et~al.}(2014){Pillepich}, {Vogelsberger}, {Deason},
  {Rodriguez-Gomez}, {Genel}, {Nelson}, {Torrey}, {Sales}, {Marinacci},
  {Springel}, {Sijacki}, \& {Hernquist}}]{pillepichetal2014}
{Pillepich}, A., {Vogelsberger}, M., {Deason}, A., {et~al.} 2014, \mnras, 444,
  237, \dodoi{10.1093/mnras/stu1408}

\bibitem[{{Rantala} {et~al.}(2018){Rantala}, {Johansson}, {Naab}, {Thomas}, \&
  {Frigo}}]{rantalaetal2018}
{Rantala}, A., {Johansson}, P.~H., {Naab}, T., {Thomas}, J., \& {Frigo}, M.
  2018, \apj, 864, 113, \dodoi{10.3847/1538-4357/aada47}

\bibitem[{{Rodriguez-Gomez} {et~al.}(2016){Rodriguez-Gomez}, {Pillepich},
  {Sales}, {Genel}, {Vogelsberger}, {Zhu}, {Wellons}, {Nelson}, {Torrey},
  {Springel}, {Ma}, \& {Hernquist}}]{rodriguez-gomezetal2016}
{Rodriguez-Gomez}, V., {Pillepich}, A., {Sales}, L.~V., {et~al.} 2016, \mnras,
  458, 2371, \dodoi{10.1093/mnras/stw456}

\bibitem[{{Salasnich} {et~al.}(2000){Salasnich}, {Girardi}, {Weiss}, \&
  {Chiosi}}]{salasnichetal2000}
{Salasnich}, B., {Girardi}, L., {Weiss}, A., \& {Chiosi}, C. 2000, \aap, 361,
  1023

\bibitem[{{Schiavon}(2007)}]{schiavon2007}
{Schiavon}, R.~P. 2007, \apjs, 171, 146, \dodoi{10.1086/511753}

\bibitem[{{Schombert}(2007)}]{schombert2007}
{Schombert}, J. 2007, ArXiv Astrophysics e-prints

\bibitem[{{Schombert}(1986)}]{schombert1986}
{Schombert}, J.~M. 1986, \apjs, 60, 603, \dodoi{10.1086/191100}

\bibitem[{{Scott} {et~al.}(2013){Scott}, {Cappellari}, {Davies}, {Kleijn},
  {Bois}, {Alatalo}, {Blitz}, {Bournaud}, {Bureau}, {Crocker}, {Davis}, {de
  Zeeuw}, {Duc}, {Emsellem}, {Khochfar}, {Krajnovi{\'c}}, {Kuntschner},
  {McDermid}, {Morganti}, {Naab}, {Oosterloo}, {Sarzi}, {Serra}, {Weijmans}, \&
  {Young}}]{scottetal2013}
{Scott}, N., {Cappellari}, M., {Davies}, R.~L., {et~al.} 2013, \mnras, 432,
  1894, \dodoi{10.1093/mnras/sts422}

\bibitem[{{Scott} {et~al.}(2017){Scott}, {Brough}, {Croom}, {Davies}, {van de
  Sande}, {Allen}, {Bland-Hawthorn}, {Bryant}, {Cortese}, {D'Eugenio},
  {Federrath}, {Ferreras}, {Goodwin}, {Groves}, {Konstantopoulos}, {Lawrence},
  {Medling}, {Moffett}, {Owers}, {Richards}, {Robotham}, {Tonini}, \&
  {Yi}}]{scottetal2017}
{Scott}, N., {Brough}, S., {Croom}, S.~M., {et~al.} 2017, \mnras, 472, 2833,
  \dodoi{10.1093/mnras/stx2166}

\bibitem[{{Skrutskie} {et~al.}(2006){Skrutskie}, {Cutri}, {Stiening},
  {Weinberg}, {Schneider}, {Carpenter}, {Beichman}, {Capps}, {Chester},
  {Elias}, {Huchra}, {Liebert}, {Lonsdale}, {Monet}, {Price}, {Seitzer},
  {Jarrett}, {Kirkpatrick}, {Gizis}, {Howard}, {Evans}, {Fowler}, {Fullmer},
  {Hurt}, {Light}, {Kopan}, {Marsh}, {McCallon}, {Tam}, {Van Dyk}, \&
  {Wheelock}}]{skrutskieetal2006}
{Skrutskie}, M.~F., {Cutri}, R.~M., {Stiening}, R., {et~al.} 2006, \aj, 131,
  1163, \dodoi{10.1086/498708}

\bibitem[{{Spinrad} \& {Taylor}(1971)}]{spinradtaylor1971}
{Spinrad}, H., \& {Taylor}, B.~J. 1971, \apjs, 22, 445, \dodoi{10.1086/190232}

\bibitem[{{Spolaor} {et~al.}(2010){Spolaor}, {Kobayashi}, {Forbes}, {Couch}, \&
  {Hau}}]{spolaoretal2010}
{Spolaor}, M., {Kobayashi}, C., {Forbes}, D.~A., {Couch}, W.~J., \& {Hau},
  G.~K.~T. 2010, \mnras, 408, 272, \dodoi{10.1111/j.1365-2966.2010.17080.x}

\bibitem[{{Thomas} {et~al.}(2005){Thomas}, {Maraston}, {Bender}, \& {Mendes de
  Oliveira}}]{thomasetal2005}
{Thomas}, D., {Maraston}, C., {Bender}, R., \& {Mendes de Oliveira}, C. 2005,
  \apj, 621, 673, \dodoi{10.1086/426932}

\bibitem[{{Thomas} {et~al.}(2010){Thomas}, {Maraston}, {Schawinski}, {Sarzi},
  \& {Silk}}]{thomasetal2010}
{Thomas}, D., {Maraston}, C., {Schawinski}, K., {Sarzi}, M., \& {Silk}, J.
  2010, \mnras, 404, 1775, \dodoi{10.1111/j.1365-2966.2010.16427.x}

\bibitem[{{Thomas} {et~al.}(2007){Thomas}, {Jesseit}, {Naab}, {Saglia},
  {Burkert}, \& {Bender}}]{thomasetal2007}
{Thomas}, J., {Jesseit}, R., {Naab}, T., {et~al.} 2007, \mnras, 381, 1672,
  \dodoi{10.1111/j.1365-2966.2007.12343.x}

\bibitem[{{Thomas} {et~al.}(2016){Thomas}, {Ma}, {McConnell}, {Greene},
  {Blakeslee}, \& {Janish}}]{thomasetal2016}
{Thomas}, J., {Ma}, C.-P., {McConnell}, N.~J., {et~al.} 2016, \nat, 532, 340,
  \dodoi{10.1038/nature17197}

\bibitem[{{Thomas} {et~al.}(2011)}]{thomasetal2011}
{Thomas}, J., {et~al.} 2011, \mnras, 415, 545,
  \dodoi{10.1111/j.1365-2966.2011.18725.x}

\bibitem[{{Trager} {et~al.}(2000{\natexlab{a}}){Trager}, {Faber}, {Worthey}, \&
  {Gonz{\'a}lez}}]{trageretal2000b}
{Trager}, S.~C., {Faber}, S.~M., {Worthey}, G., \& {Gonz{\'a}lez}, J.~J.
  2000{\natexlab{a}}, \aj, 120, 165, \dodoi{10.1086/301442}

\bibitem[{{Trager} {et~al.}(2000{\natexlab{b}}){Trager}, {Faber}, {Worthey}, \&
  {Gonz{\'a}lez}}]{trageretal2000a}
---. 2000{\natexlab{b}}, \aj, 119, 1645, \dodoi{10.1086/301299}

\bibitem[{{Trager} {et~al.}(1998){Trager}, {Worthey}, {Faber}, {Burstein}, \&
  {Gonz{\'a}lez}}]{trageretal1998}
{Trager}, S.~C., {Worthey}, G., {Faber}, S.~M., {Burstein}, D., \&
  {Gonz{\'a}lez}, J.~J. 1998, \apjs, 116, 1, \dodoi{10.1086/313099}

\bibitem[{{Tran} {et~al.}(2010){Tran}, {Papovich}, {Saintonge}, {Brodwin},
  {Dunlop}, {Farrah}, {Finkelstein}, {Finkelstein}, {Lotz}, {McLure},
  {Momcheva}, \& {Willmer}}]{tranetal2010}
{Tran}, K.-V.~H., {Papovich}, C., {Saintonge}, A., {et~al.} 2010, \apjl, 719,
  L126, \dodoi{10.1088/2041-8205/719/2/L126}

\bibitem[{{Valentinuzzi} {et~al.}(2010){Valentinuzzi}, {Poggianti}, {Saglia},
  {Arag{\'o}n-Salamanca}, {Simard}, {S{\'a}nchez-Bl{\'a}zquez}, {D'onofrio},
  {Cava}, {Couch}, {Fritz}, {Moretti}, \& {Vulcani}}]{valentinuzzietal2010}
{Valentinuzzi}, T., {Poggianti}, B.~M., {Saglia}, R.~P., {et~al.} 2010, \apjl,
  721, L19, \dodoi{10.1088/2041-8205/721/1/L19}

\bibitem[{{van de Sande} {et~al.}(2018){van de Sande}, {Scott},
  {Bland-Hawthorn}, {Brough}, {Bryant}, {Colless}, {Cortese}, {Croom},
  {d'Eugenio}, {Foster}, {Goodwin}, {Konstantopoulos}, {Lawrence}, {McDermid},
  {Medling}, {Owers}, {Richards}, \& {Sharp}}]{vandesandeetal2018}
{van de Sande}, J., {Scott}, N., {Bland-Hawthorn}, J., {et~al.} 2018, Nature
  Astronomy, 2, 483, \dodoi{10.1038/s41550-018-0436-x}

\bibitem[{{van der Wel} {et~al.}(2008){van der Wel}, {Holden}, {Zirm}, {Franx},
  {Rettura}, {Illingworth}, \& {Ford}}]{vanderweletal2008}
{van der Wel}, A., {Holden}, B.~P., {Zirm}, A.~W., {et~al.} 2008, \apj, 688,
  48, \dodoi{10.1086/592267}

\bibitem[{{van der Wel} {et~al.}(2014){van der Wel}, {Franx}, {van Dokkum},
  {Skelton}, {Momcheva}, {Whitaker}, {Brammer}, {Bell}, {Rix}, {Wuyts},
  {Ferguson}, {Holden}, {Barro}, {Koekemoer}, {Chang}, {McGrath},
  {H{\"a}ussler}, {Dekel}, {Behroozi}, {Fumagalli}, {Leja}, {Lundgren},
  {Maseda}, {Nelson}, {Wake}, {Patel}, {Labb{\'e}}, {Faber}, {Grogin}, \&
  {Kocevski}}]{vanderweletal2014}
{van der Wel}, A., {Franx}, M., {van Dokkum}, P.~G., {et~al.} 2014, \apj, 788,
  28, \dodoi{10.1088/0004-637X/788/1/28}

\bibitem[{{van Dokkum} {et~al.}(2008){van Dokkum}, {Franx}, {Kriek}, {Holden},
  {Illingworth}, {Magee}, {Bouwens}, {Marchesini}, {Quadri}, {Rudnick},
  {Taylor}, \& {Toft}}]{vandokkumetal2008}
{van Dokkum}, P.~G., {Franx}, M., {Kriek}, M., {et~al.} 2008, \apjl, 677, L5,
  \dodoi{10.1086/587874}

\bibitem[{{Veale} {et~al.}(2017{\natexlab{a}}){Veale}, {Ma}, {Greene},
  {Thomas}, {Blakeslee}, {McConnell}, {Walsh}, \& {Ito}}]{vealeetal2017b}
{Veale}, M., {Ma}, C.-P., {Greene}, J.~E., {et~al.} 2017{\natexlab{a}}, \mnras,
  471, 1428, \dodoi{10.1093/mnras/stx1639}

\bibitem[{{Veale} {et~al.}(2018){Veale}, {Ma}, {Greene}, {Thomas}, {Blakeslee},
  {Walsh}, \& {Ito}}]{vealeetal2018}
---. 2018, \mnras, 473, 5446, \dodoi{10.1093/mnras/stx2717}

\bibitem[{{Veale} {et~al.}(2017{\natexlab{b}}){Veale}, {Ma}, {Thomas},
  {Greene}, {McConnell}, {Walsh}, {Ito}, {Blakeslee}, \&
  {Janish}}]{vealeetal2017a}
{Veale}, M., {Ma}, C.-P., {Thomas}, J., {et~al.} 2017{\natexlab{b}}, \mnras,
  464, 356, \dodoi{10.1093/mnras/stw2330}

\bibitem[{{Vogelsberger} {et~al.}(2013){Vogelsberger}, {Genel}, {Sijacki},
  {Torrey}, {Springel}, \& {Hernquist}}]{vogelsbergeretal2013}
{Vogelsberger}, M., {Genel}, S., {Sijacki}, D., {et~al.} 2013, \mnras, 436,
  3031, \dodoi{10.1093/mnras/stt1789}

\bibitem[{{Wake} {et~al.}(2012){Wake}, {van Dokkum}, \& {Franx}}]{wakeetal2012}
{Wake}, D.~A., {van Dokkum}, P.~G., \& {Franx}, M. 2012, \apjl, 751, L44,
  \dodoi{10.1088/2041-8205/751/2/L44}

\bibitem[{{Wellons} {et~al.}(2016){Wellons}, {Torrey}, {Ma}, {Rodriguez-Gomez},
  {Pillepich}, {Nelson}, {Genel}, {Vogelsberger}, \&
  {Hernquist}}]{wellonsetal2016}
{Wellons}, S., {Torrey}, P., {Ma}, C.-P., {et~al.} 2016, \mnras, 456, 1030,
  \dodoi{10.1093/mnras/stv2738}

\bibitem[{{Wetzel} {et~al.}(2013){Wetzel}, {Tinker}, {Conroy}, \& {van den
  Bosch}}]{wetzeletal2013}
{Wetzel}, A.~R., {Tinker}, J.~L., {Conroy}, C., \& {van den Bosch}, F.~C. 2013,
  \mnras, 432, 336, \dodoi{10.1093/mnras/stt469}

\bibitem[{{White}(1980)}]{white1980}
{White}, S.~D.~M. 1980, \mnras, 191, 1P, \dodoi{10.1093/mnras/191.1.1P}

\bibitem[{{Woo} {et~al.}(2013){Woo}, {Dekel}, {Faber}, {Noeske}, {Koo},
  {Gerke}, {Cooper}, {Salim}, {Dutton}, {Newman}, {Weiner}, {Bundy}, {Willmer},
  {Davis}, \& {Yan}}]{wooetal2013}
{Woo}, J., {Dekel}, A., {Faber}, S.~M., {et~al.} 2013, \mnras, 428, 3306,
  \dodoi{10.1093/mnras/sts274}

\bibitem[{{Worthey} {et~al.}(1992){Worthey}, {Faber}, \&
  {Gonzalez}}]{wortheyetal1992}
{Worthey}, G., {Faber}, S.~M., \& {Gonzalez}, J.~J. 1992, \apj, 398, 69,
  \dodoi{10.1086/171836}

\bibitem[{{Worthey} {et~al.}(2014){Worthey}, {Tang}, \&
  {Serven}}]{wortheyetal2014}
{Worthey}, G., {Tang}, B., \& {Serven}, J. 2014, \apj, 783, 20,
  \dodoi{10.1088/0004-637X/783/1/20}

\bibitem[{{Wu} {et~al.}(2014){Wu}, {Gerhard}, {Naab}, {Oser},
  {Martinez-Valpuesta}, {Hilz}, {Churazov}, \& {Lyskova}}]{wuetal2014}
{Wu}, X., {Gerhard}, O., {Naab}, T., {et~al.} 2014, \mnras, 438, 2701,
  \dodoi{10.1093/mnras/stt2415}

\end{thebibliography}

\end{document}